\documentclass[a4paper,fleqn,usenatbib]{mnras}
\usepackage{newtxtext,newtxmath}

\usepackage[T1]{fontenc}
\usepackage{ae,aecompl}
\usepackage{hyperref}

\usepackage{graphicx}   
\usepackage{amsmath}    
\usepackage{amssymb}    
\usepackage{subfigure}
\usepackage{float}

\title[Investigation of RRATs]{A Long-term study of three rotating radio transients}

\author[B. Bhattacharyya  et al.]{
B.~Bhattacharyya,$^{1}$\thanks{E-mail: bhaswati@ncra.tifr.res.in}
A.~G.~Lyne,$^{2}$ B.~W.~Stappers,$^{2}$ P.~Weltevrede,$^{2}$ E.~F.~Keane,$^{2,3}$ \and M.~A.~McLaughlin,$^{4,5}$ M. Kramer,$^{2,6}$ C. Jordan,$^{2}$ C. Bassa$^{7}$
\\
$^{1}$ National Centre for Radio Astrophysics, Tata Institute of Fundamental Research, Pune 411 007, India \\
$^{2}$ Jodrell Bank Centre for Astrophysics, School of Physics and Astronomy, The University of Manchester, Manchester M13 9PL, UK\\ 
$^{3}$ SKA Organisation, Jodrell Bank Observatory, SK11 9DL, UK\\
$^{4}$ Department of Physics \& Astronomy, West Virginia University, Morgantown, WV 26506, US\\
$^{5}$ Center for Gravitational Waves and Cosmology, West Virginia University, Chestnut Ridge Research Building, Morgantown, WV 26505\\
$^{6}$ Max-Planck-Institut fu ̈r Radioastronomie, Auf dem Hu ̈gel 69, D-53121 Bonn, Germany\\
$^{7}$ ASTRON, the Netherlands Institute for Radio Astronomy, Postbus 2, 7990 AA, Dwingeloo, The Netherlands\\
}

\date{Accepted XXX. Received YYY; in original form ZZZ}

\pubyear{2017}

\begin{document}
\label{firstpage}
\pagerange{\pageref{firstpage}--\pageref{lastpage}}
\maketitle

\begin{abstract}
We present the longest-term timing study so far of three Rotating Radio Transients (RRATs) $-$ J1819$-$1458, J1840$-$1419 and J1913$+$1330 $-$ 
performed using the Lovell, Parkes and Green Bank telescopes over the past decade.  We study long-term and short-term variations of the pulse 
emission rate from these RRATs and report a marginal indication of a long-term increase in pulse detection rate over time for PSR J1819$-$1458 
and J1913$+$1330. For PSR J1913$+$1330, we also observe a two orders of magnitude variation in the observed pulse detection
rates across individual epochs, which may constrain the models explaining the origin of RRAT pulses. PSR J1913$+$1330 is also observed 
to exhibit a weak persistent emission mode. 

We investigate the post-glitch timing properties of J1819$-$1458 (the only RRAT for which glitches are observed) and discuss the implications 
for possible glitch models. Its post-glitch over-recovery of the frequency derivative is magnetar-like and similar behaviour is only observed 
for two other pulsars, both of which have relatively high magnetic field strengths. Following the over-recovery we also observe that some 
fraction of the pre-glitch frequency derivative is gradually recovered. 
\end{abstract}





\section{Introduction}                \label{sec:intro}  
Occasional flashes of dispersed radio emission of typically a few milliseconds duration are detected from the 
Rotating Radio Transients (RRATs; \cite{mclaughlin06}). Even though the pulses appear randomly, there is a characteristic 
underlying periodicity associated with the emission detected from the RRATs. 
Timing studies and multi-wavelength observations have revealed that RRATs are neutron stars, most likely an extreme 
manifestation of the overall neutron star intermittency spectrum. A decade since their discovery by \cite{mclaughlin06}, 
there are 112 known RRATs\footnote{http://astro.phys.wvu.edu/rratalog/} having spin periods ranging from 0.125 s to 7.7 s and 
dispersion measures ranging from 9.2 pc cm$^{-3}$ to 554.9 pc cm$^{-3}$. Measurements of the period derivative exist for 
only 29 of RRATs, ranging from 5.7$\times$10$^{-13}$ to 1.2$\times$10$^{-16}$ s/s. The period and magnetic field strength 
distributions of the RRATs are skewed to larger values compared to that of the normal pulsars, with some comparable to 
those of X-ray detected radio-quiet isolated neutron stars and magnetars \citep{cui17}. 
The origin of RRAT emission is not yet known and a number of postulates exist in the literature. To name a few, the pulses 
observed from RRATs are thought to be associated with, (a) giant pulses from weak pulsars \citep{knight06},                
(b) a manifestation of extreme nulling of radio pulsars \citep{redman09}, (c) created due to the presence of a
circumstellar asteroid/radiation belt around the pulsar \citep{cordes08}, or (d) from systems similar
to PSR B0656$+$14, for which emission properties would have been similar to the RRATs if this pulsar is placed at a
larger distance \citep{weltevrede06}. Therefore we do not know if RRATs represent a truly separate population of radio emitting neutron 
stars like magnetars or isolated neutron stars. Phase-connected timing solutions for RRATs provide timing models with information about the 
period, period derivative, magnetic field strength and spin down energy rate; enabling us to compare these properties with rest of the 
neutron star population. Timing solutions are also important to obtain accurate positions which facilitate identification of possible high energy
counterparts. 
 
In this paper we present results from long-term monitoring of three RRATs, J1819$-$1458, J1840$-$1419 (originally known as J1841$-$1418) 
and J1913$+$1330 (originally known as J1913$+$1333). This study reports results for regular observations of these RRATs over 
the past decade and presents the longest time-span investigation of RRATs.
 
The brightest known RRAT is J1819$-$1458. This is one of the first RRATs discovered by \cite{mclaughlin06}. It has a wide multi-component  
profile and is located in the upper right part of the $P-\dot P$ diagram, in the same area occupied by the magnetars and high 
magnetic field pulsars. PSR J1819$-$1458 is the only RRAT that is also detected in X$-$rays \citep{rea09}. The detection of an 
X$-$ray counterpart with properties similar to those of other neutron stars provides a strong link to relate RRATs with the greater neutron star 
population.  Extended X-ray emission is also detected around PSR J1819$-$1458 \citep{camero13}, which can be interpreted as being a nebula 
powered by the RRAT. \cite{dhillon11} attempted to detect optical emission from simultaneous ULTRACAM on the William Herschel 
Telescope and Lovell observations, and found no evidence of optical pulses at magnitudes brighter than $i$=19.3 to a 5$\sigma$ limit.
\cite{karastergiou09} studied the polarisation properties of J1819$-$1458 with Parkes at 1420 MHz. The polarisation 
characteristics and integrated profile resemble those of normal pulsars with average spin-down energy $\dot{E}$, and 
a smooth S-shaped swing of polarisation position angle. \cite{lyne09} 
presented a timing analysis of PSR J1819$-$1458 starting from the discovery observations on 1998 August, 
followed by 5 years of timing starting in 2003. They reported the detection of two glitches (characterised by sudden 
jumps in rotational frequency) from this RRAT, having similar magnitude to the glitches observed for radio pulsars and 
magnetars. So far it is the only RRAT for which glitches are observed. The lack of glitches observed in RRATs can be explained 
by the fact that they appear to represent a slightly older population of the neutron stars and glitches are generally observed in younger pulsars. 
Moreover, very few RRATs have timing solutions or even being regularly timed. \cite{lyne09} observed atypical post-glitch properties 
for PSR J1819$-$1458. 
The glitches resulted in a long term reduction in the average spin-down rate as opposed to the increase of average spin-down rate 
generally observed for pulsars.

PSR J1840$-$1419 was discovered in a re-analysis of the Parkes multi-beam survey \citep{keane10}. 
\cite{keane11} reported its coherent timing solution for the data span from March 2009 to October 2010.
\cite{keane13} performed X$-$ray observations of J1840$-$1419 and calculated a 
blackbody temperature upper limit, implying that this RRAT is one of the coolest neutron stars known. 
 
PSR J1913$+$1330 was discovered by \cite{mclaughlin06}. The timing solution of PSR J1913$+$1330 from 
January 2004 to April 2009 was presented by \cite{mclaughlin09}. They observed that PSR J1913$+$1330 has spin properties indistinguishable 
from the rest of the radio pulsar population. 

In \S \ref{sec:obs} we describe the observations. In \S \ref{sec:burst_rate} we report
an investigation of the pulse rate statistics of these three RRATs. \S \ref{sec:timing_J1819} describes the timing study of PSR J1819$-$1458. 
\S \ref{sec:timing_J1840} and \S \ref{sec:timing_J1913} details the timing study of PSR J1840$-$1419 and PSR J1913$+$1330 respectively. 
\S \ref{sec:average_det_J1913} presents the detection of a weak emission mode for PSR J1913$+$1330. In \S \ref{sec:discussion} 
we discuss and summarise the main results from this study. 

\begin{figure*}
\centering
\subfigure[PSR J1819$-$1458]{
\includegraphics[width=6in]{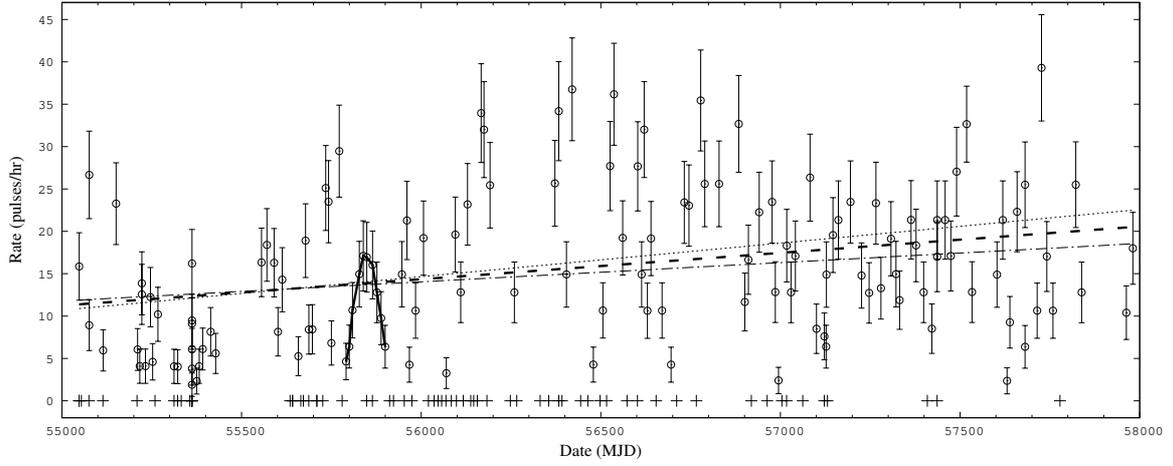}
\label{burst_1819}
}\\
\subfigure[PSR J1840$-$1419]{
\includegraphics[width=6in]{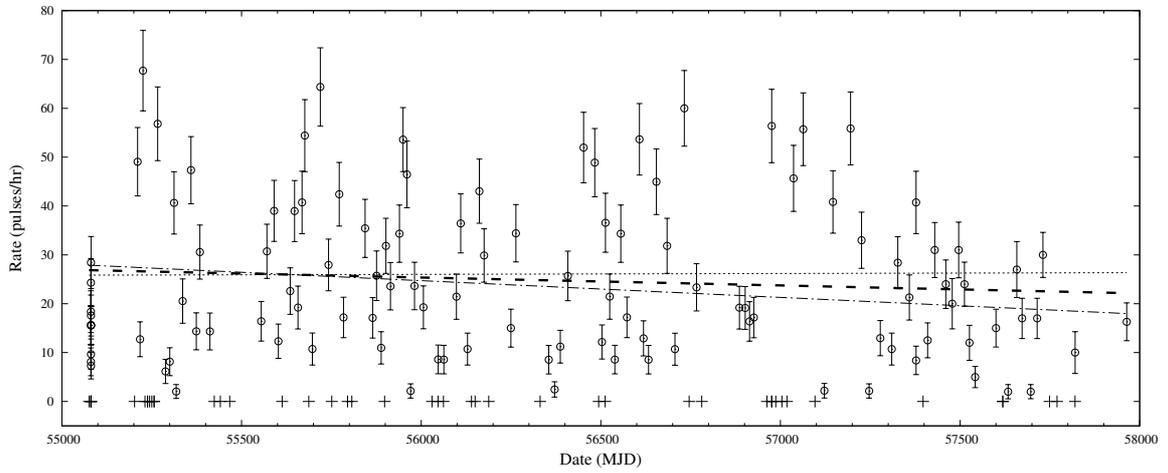}
\label{burst_1841}
}\\
\subfigure[PSR J1913$+$1330]{
\includegraphics[width=6in]{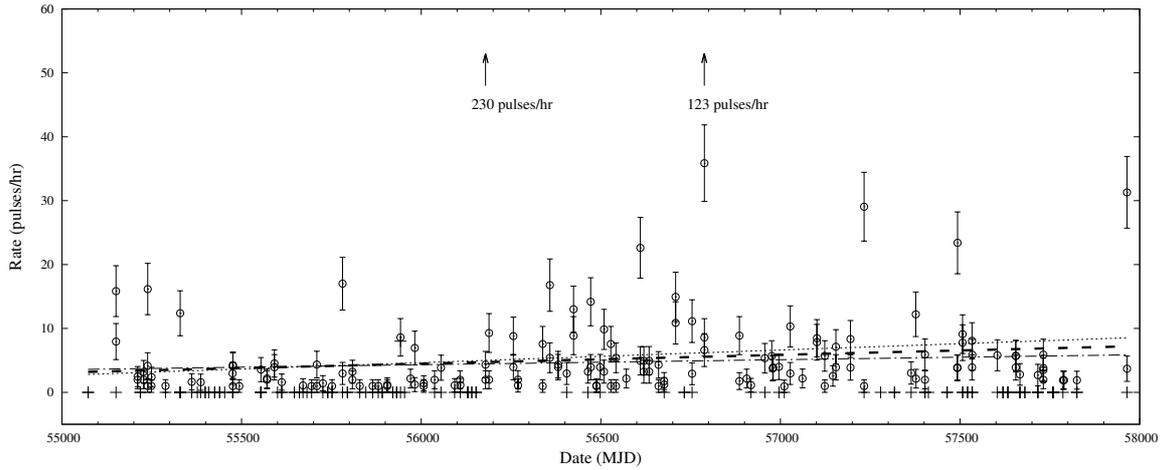}
\label{burst_1913}
}
\caption {Pulse detection rate at each observing epoch vs the MJD of that epoch, for the RRATs J1819$-$1458, J1840$-$1419 and J1913$+$1330. Statistical error bars for each data rate are plotted \citep{gehrels86}. The thick dashed lines are linear fits to the data, while the dotted line and the dash-dot lines represent the 1$\sigma$ error on the fitted line. Parameters for the fitted straight lines are presented in Table \ref{tab:fig1_burst_stat}. In all the cases the reduced chi-square is much more than unity, indicating inconsistency with a fixed emission rate for these RRATs from day to day. Two epochs with order of magnitude higher detection rates for PSR J1913$+$1330 are denoted by the upwards arrows. The epochs where no pulses were detected from the RRATs are indicated by `+'.}
\label{rate}
\end{figure*}

\section{Observations and analysis}  \label{sec:obs}
The observations were carried out using the 64-m Parkes Telescope in Australia and 76-m the Lovell telescope at Jodrell Bank in the UK at  
frequencies of around 1.4 GHz and the 100-m Green Bank Telescope in the USA at 2.2 GHz. The observations up to March 2009 were reported 
in \cite{lyne09} and \cite{mclaughlin09}. Building on these previously reported results, we have observed these pulsars for at least 8 more years 
with the 76-m Lovell telescope. At the Parkes telescope, dual orthogonal linear polarisations were added to generate total intensity 
recorded after forming a 512$\times$0.5 MHz filter bank, with sampling resolution of 100 $\upmu$s. At the Lovell telescope dual 
orthogonal circular polarisations were added to generate total intensity.
Observations between March 2009 and August 2009 were performed with the analog filterbank backend (AFB) 
with 64 MHz bandwidth with 100 $\upmu$s time resolution. Observations after August 2009 till May 2016 were recorded 
with the digital backend (DFB) \citep{hobbs04}, with 300 MHz bandwidth and 100 $\upmu$s time resolution. 
Because of the increased bandwidth, the sensitivity of the DFB backend is $\sim$2 times greater than that of the AFB backend. 
The observations were mostly of 30 mins in duration. The data were affected by radio frequency interference (RFI). We masked a 
standard list of RFI frequency channels for the Lovell telescope coming from known RFI sources, and removed other RFI occurences 
by visual inspection. 

Pulsar timing is normally performed with times-of-arrival (TOAs) calculated from integrated pulse profiles generated by adding a large number 
of (typically $\sim$1000) single pulses folded with the known pulsar period, to provide increased signal-to-noise and a stable pulse 
profile. For timing of RRATs, we need to work with individual pulses as opposed to the integrated profiles because of the 
sporadic nature of their emission. The detected single pulses from RRATs are generally quite strong, with typical peak flux densities 
of $\sim$10$^2$$-$10$^3$ mJy \citep{keane11}. For the three RRATs studied in this paper, the signal-to-noise ratio of the individual 
pulses is sufficient to generate TOAs from each pulse.
For this purpose we dedispersed the data with a range of dispersion measure (DM) values around the DM of the RRAT and also at a 
DM of zero. Then we searched for pulses above 5$\sigma$ from both the time series using the {\sc sigproc}\footnote{http://sigproc.sourceforge.net} 
pulsar data processing package. Results from both the searches are compared and those pulses with stronger detection at the DM 
of the RRAT than at a DM of zero were considered. We improved the data quality by using zero-DM filtering \citep{eatough09}. Finally 
visual investigation of detections was performed, eliminating pulses that are outside the expected pulse window, which are likely 
generated from sources of interference.
This allowed us to study the burst rate and its evolution with time as detailed in \S \ref{sec:burst_rate}. 
Then the barrycentric TOAs for each single pulse are calculated by correlating it with a template of a strong pulse of the RRAT. As 
individual pulses are typically narrower, relatively broader templates based on the average profiles will not be suitable for correlating with 
the single pulses. The TOAs are modeled using the standard pulsar timing software {\sc tempo}\footnote{\url{tempo.sourceforge.net}}, following exactly the same method as for normal pulsars \citep{lorimer}.

\begin{figure}
\begin{minipage}{\columnwidth}
\includegraphics[width=3.0in]{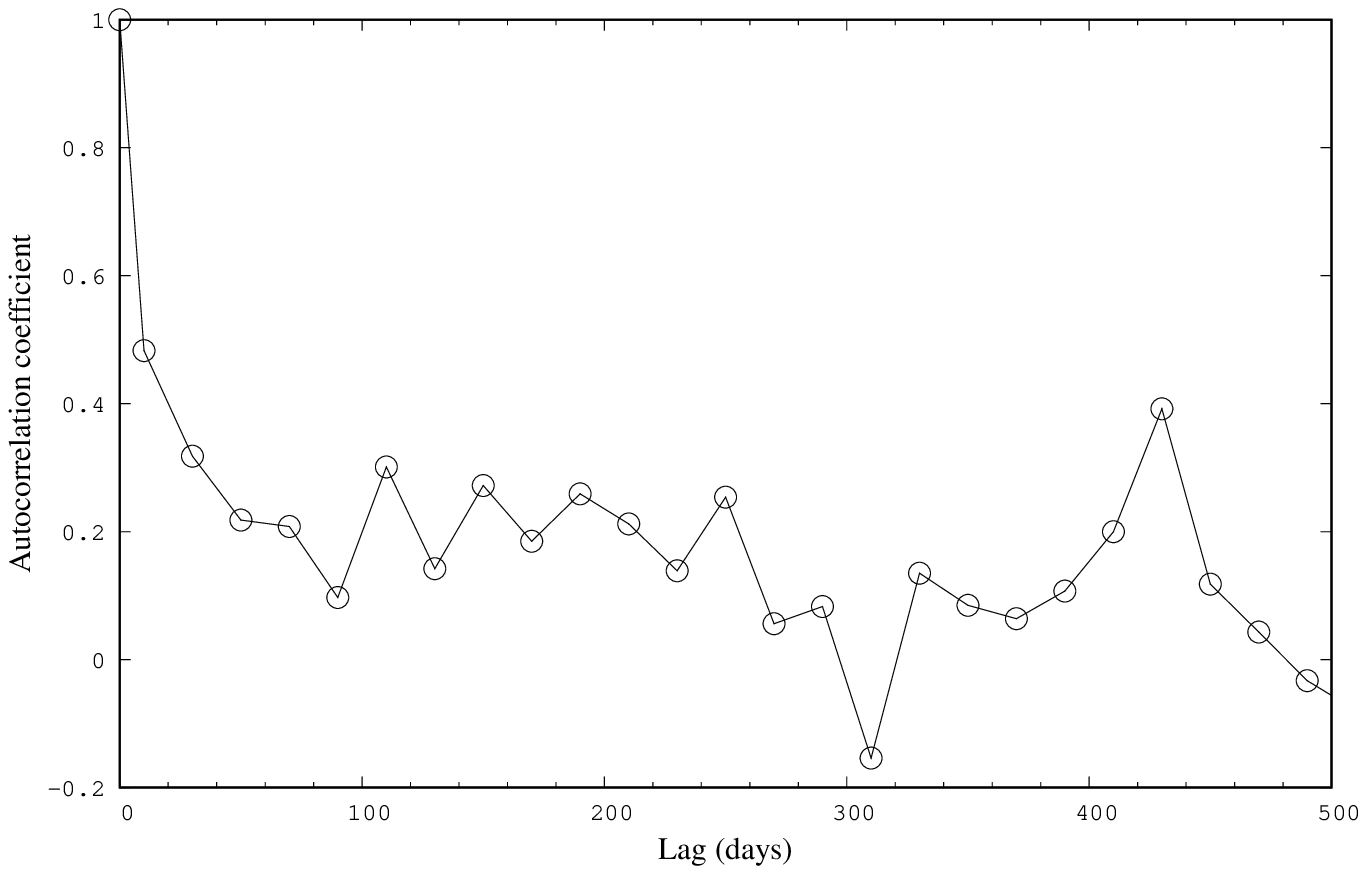}
\end{minipage}
\begin{minipage}{\columnwidth}
\includegraphics[width=3.0in]{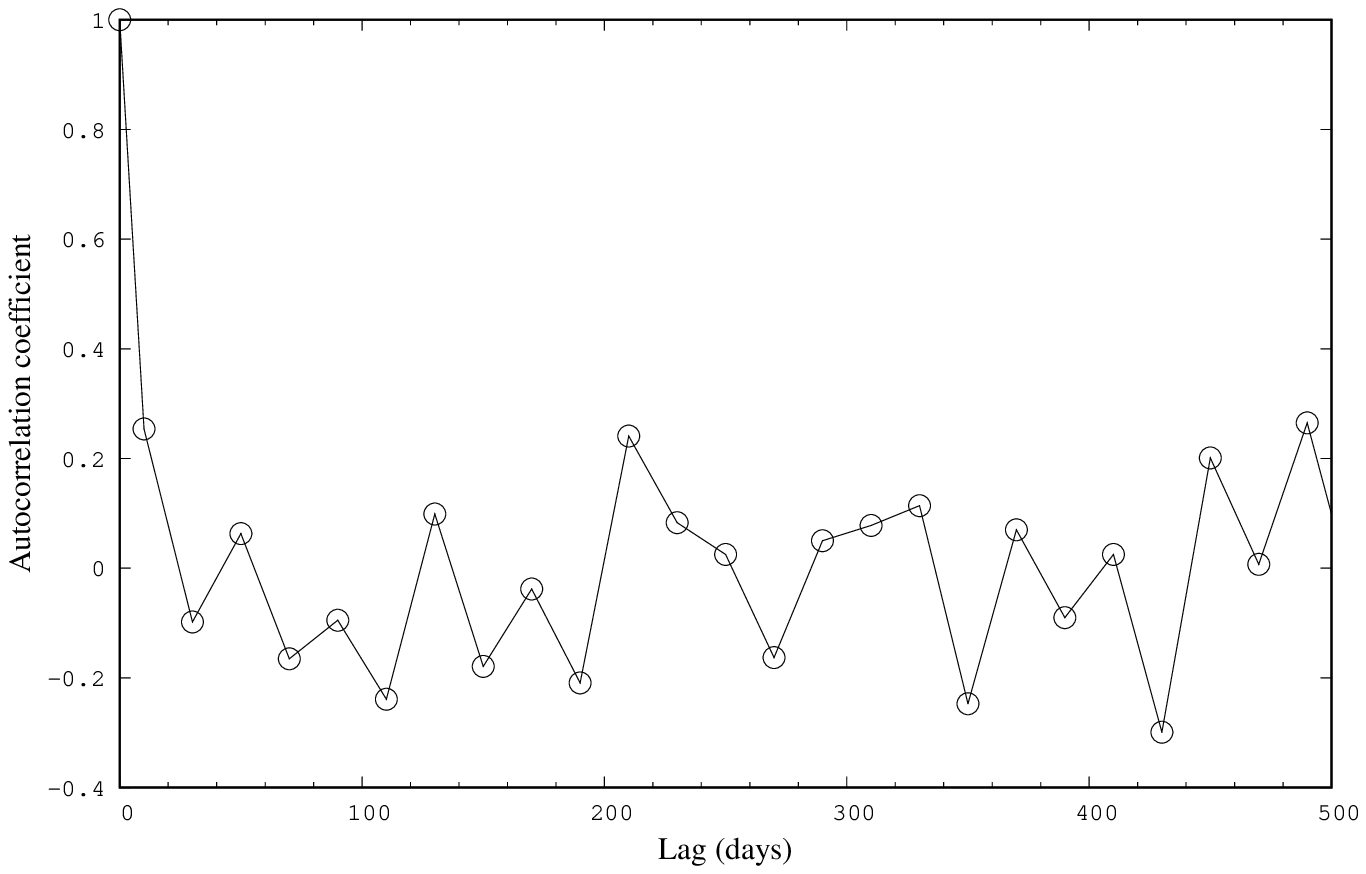}
\end{minipage}
\begin{minipage}{\columnwidth}
\includegraphics[width=3.0in]{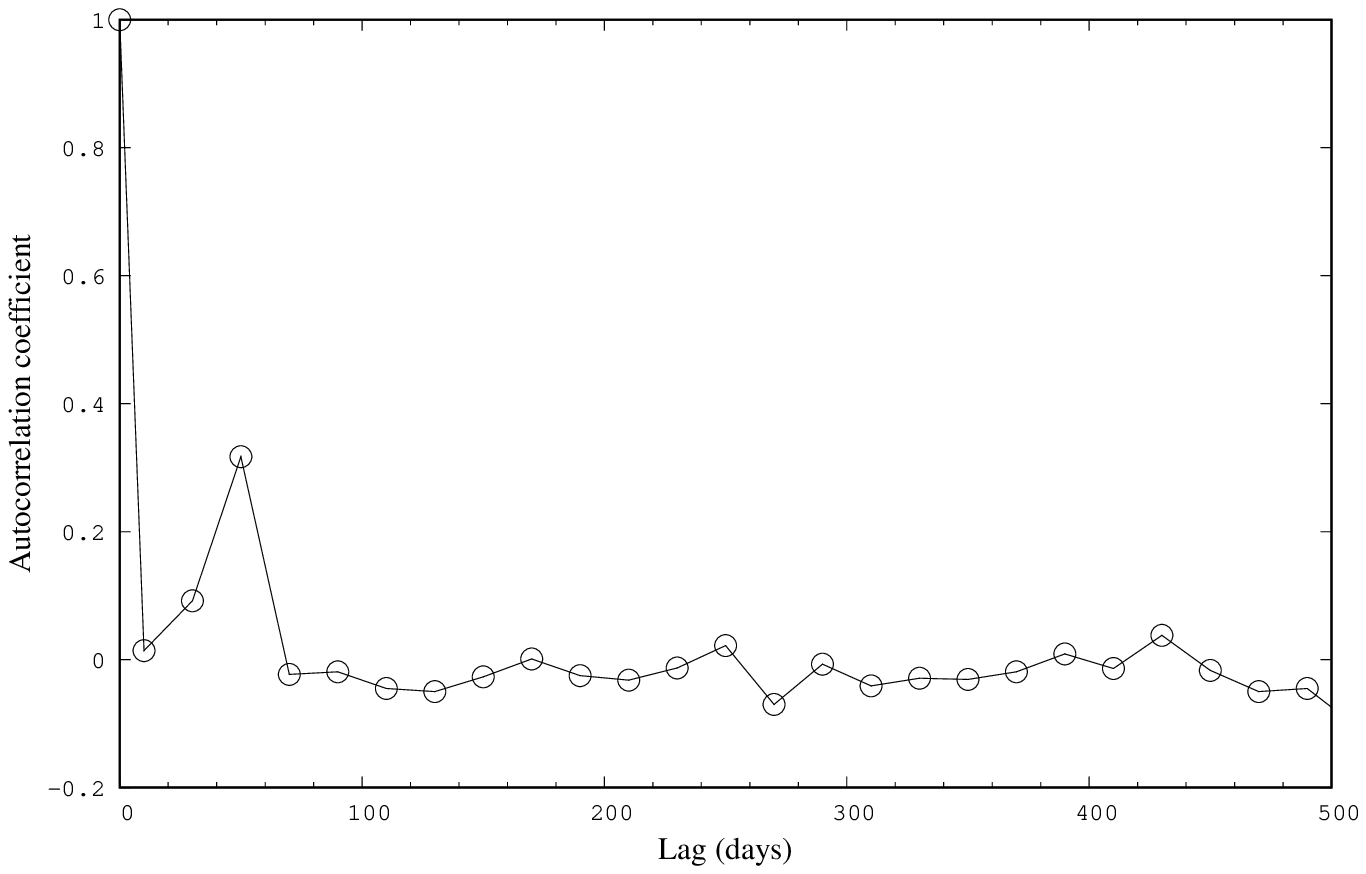}
\end{minipage}
\caption{Autocorrelation analysis of the pulse detection rate variation shown in Figure \ref{rate}, for PSR J1819$-$1458 (top panel), for PSR J1840$-$1419 (middle panel) and for PSR J1913$-$1330 (bottom panel).} 
\label{fig:autocor}
\end{figure}

\begin{figure}
\begin{center}
\includegraphics[angle=0,width=0.5\textwidth]{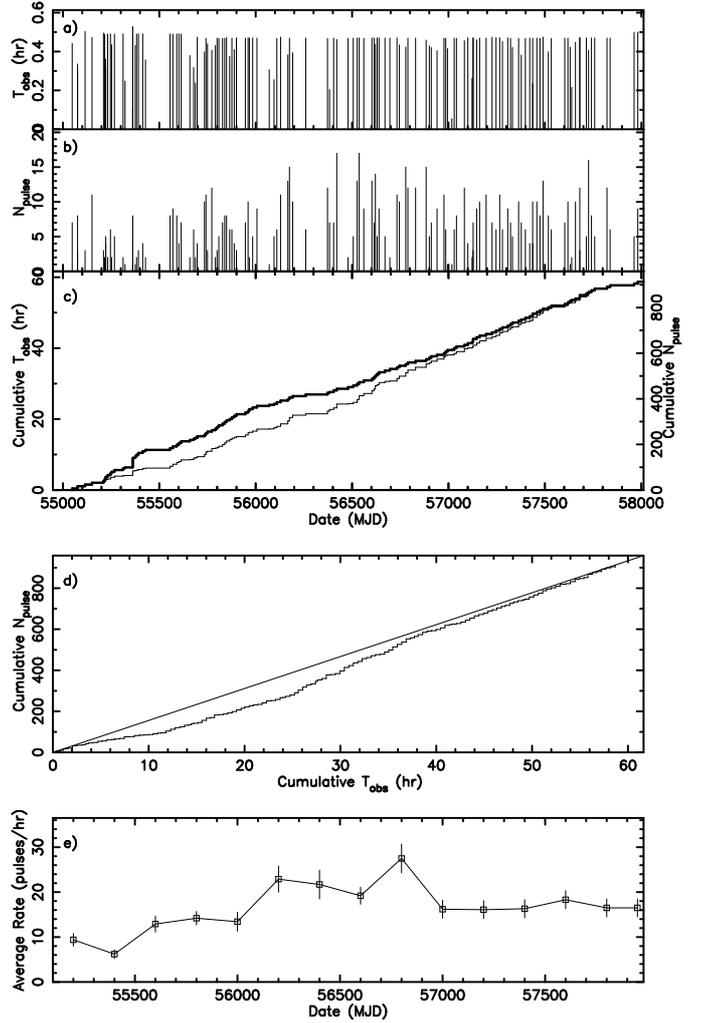}
\caption[tobservation and number of pulses J1819$-$1458]{Statistics of pulses detected from PSR J1819$-$1458. (a) Duration of observation (T$_{\text{obs}}$) vs the epoch of observation (in MJD) in which the RRAT was detected, (b) Number of pulses (N$_{\text{pulse}}$) detected vs the epoch of observation (MJD), (c) Cumulative T$_{\text{obs}}$ (with thin line) and Cumulative N$_{\text{pulse}}$ (with heavy line) vs the epoch of observation, (d) Cumulative T$_{\text{obs}}$ vs Cumulative N$_{\text{pulse}}$, solid line represents the mean burst rate over the range of observations shown, (e) Average rate (pulses/hr) vs the epoch (MJD).}
\label{tobs_npulse_1819}
\vspace{0.1cm}
\end{center}
\end{figure}
\begin{figure}
\begin{center}
\includegraphics[angle=0,width=0.5\textwidth]{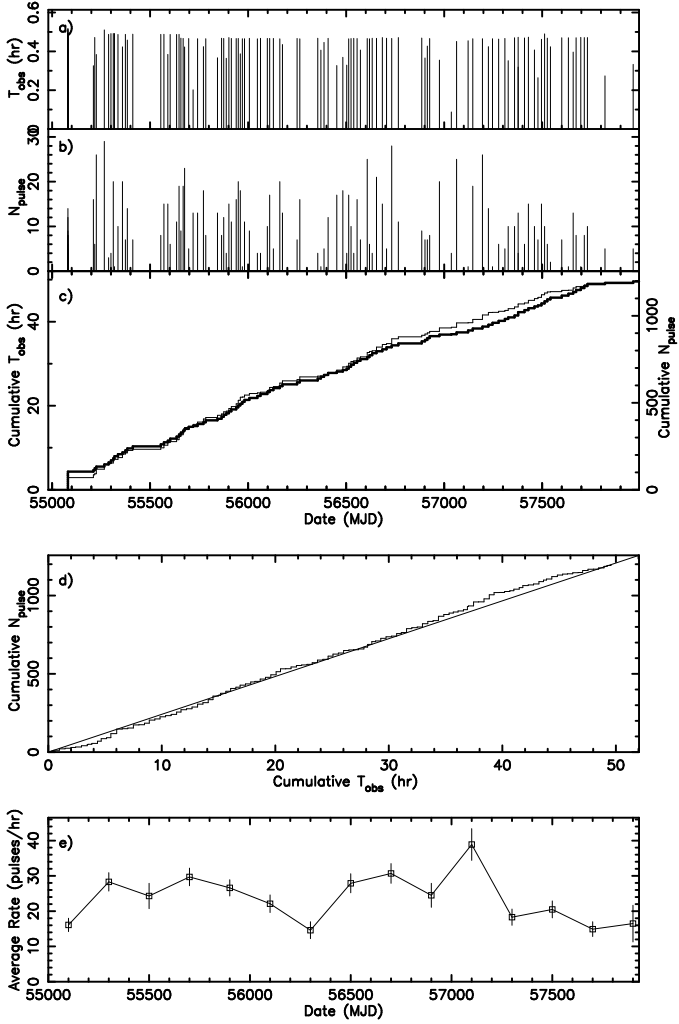}
\caption[tobservation and number of pulses J1840$-$1419]{Statistics of pulses detected from PSR J1840$-$1419. The panels are as described for Figure \ref{tobs_npulse_1819}.}
\label{tobs_npulse_1841}
\vspace{0.1cm}
\end{center}
\end{figure}
\begin{figure}
\begin{center}
\includegraphics[angle=0,width=0.5\textwidth]{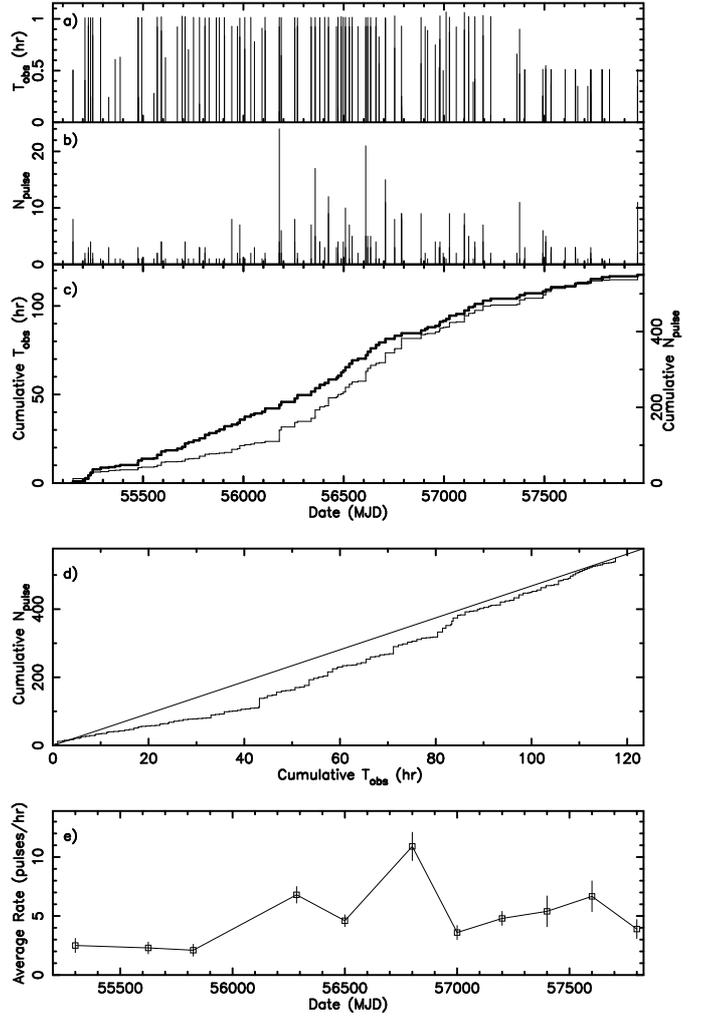}
\caption[tobservation and number of pulses J1913$+$1330]{Statistics of pulses detected from PSR J1913$+$1330. The panels are as described for Figure \ref{tobs_npulse_1819}.}
\label{tobs_npulse_1913}
\vspace{0.1cm}
\end{center}
\end{figure}
\begin{table*}
\caption{Parameters for the straight line fitting, presented in Figure \ref{rate}.}
\label{tab:fig1_burst_stat}
\vspace{0.3cm}
\begin{tabular}{|l|c|c|c|c|c|c|c|c|c|c|c|c|c|c|c|c}
\hline
  RRAT name    & Slope           & Intercept$\dagger$                      & Reduced  \\
               &                 & (pulses/hr)                    & chi-square \\
               &                 &                                &           \\
  J1819$-$1458 & 0.004$\pm$0.001 & 15.9$\pm$0.7                   &   70\\
  J1840$-$1419 & $-$0.002$\pm$0.002 & 24.5$\pm$1.7                   &   279\\
  J1913$+$1330 & 0.0013$\pm$0.0005 & 4.9$\pm$0.5                   &   32\\\hline
\end{tabular}
\\
$\dagger$ Epoch for the intercept is at MJD 56500\\
\end{table*}

\section{Results}  \label{sec:results}
\subsection{Pulse rate statistics} \label{sec:burst_rate}
The RRATs were not detected in all observing epochs, in some cases because of RFI. PSR J1819$-$1458 was detected in 132 observing 
epochs out of 200 with a maximum detection rate of $\sim$38 pulses/hour. 
PSR J1840$-$1419 was detected in 114 epochs out of 160 with a maximum detection rate of $\sim$69 pulses/hour. PSR J1913$+$1330 was detected in 130 
epochs out of 210 with a maximum detection rate of $\sim$230 pulses/hour. 
We note that the duration of the observing epoch with detection rate of $\sim$230 pulses/hour is relatively short ($\sim$6 mins) compared 
to the typical observing epochs ($\sim$30 mins). This indicates that emission rates from RRATs can apparently reach a high value for 
short periods of time. 
Figure \ref{rate} shows the variation of the rate of pulse detection per hour for PSR J1819$-$1458, PSR J1840$-$1419 and PSR J1913$+$1330 based on 
the observations with the Lovell telescope. The rate of detection of the pulses varies greatly for all the three RRATs. We have fitted a 
linear relation to the detection rate versus date (MJD) data of these three RRATs. 
The slope and intercept of best fit straight lines for RRATs J1819$-$1458, J1840$-$1419 and J1913$+$1330 are presented in Table \ref{tab:fig1_burst_stat}. 
The reduced chi-square value for each fit is very high ($\gg$ 1), indicating highly variable detection rates from day to day. The fitting indicate that there is a possibility of long-term increase in the detection rate for PSR J1819$-$1458. For PSR
J1913$+$1330 there is marginal evidence of long-term increase in the detection rate; whereas no long-term change in the emission rate 
is observed for PSR J1840$-$1419. 

For PSR J1819$-$1458 we observe an instance of apparently correlated change in emission rate, as seen from MJD 55790 to MJD 55900 
(data points joined by a solid line in top panel of Figure \ref{rate}). There is some evidence of other periods where such features may have occurred but the cadence does 
not completely sample them. 
To investigate these possible correlations between pulse emission rates in nearby epochs, we conducted an autocorrelation analysis of the 
time sequence of Figure \ref{rate}. The autocorrelations for RRATs J1819$-$1458, J1840$-$1419 and J1913$+$1330 calculated for lags up to 500 days 
with 20 days of resolution are presented in Figures \ref{fig:autocor}. For PSR J1819$-$1458 
there is some correlation for lags up to about 50 days. This is consistent with the structures seen in Figure \ref{burst_1819}. 
For PSR J1840$-$1458 and PSR J1913$+$1330 the autocorrelation function falls rather fast with increasing lag values. However, for PSR  
J1913$+$1330 a significant secondary peak is observed at a lag of 50 days.

To further investigate the pulse rate statistics we have plotted the cumulative duration of observation against the epoch of 
observation as was performed for two intermittent pulsars by \cite{lyne16}. Figures \ref{tobs_npulse_1819}, \ref{tobs_npulse_1841} 
and \ref{tobs_npulse_1913} present the pulse rate statistics for RRATs J1819$-$1458, J1840$-$1419 and J1913$+$1330 respectively. 
In these diagrams, panel (a) shows the duration of 
observation (T$_{\text{obs}}$) vs the epoch of observation (in MJD). The number of detected pulses (N$_{\text{pulse}}$) vs the 
epoch of observation (MJD) is plotted in panel (b). Panel (c) shows the cumulative T$_{\text{obs}}$ (with thin line) and 
the cumulative N$_{\text{pulse}}$ (with heavy line) versus the epoch of observation, whereas the cumulative T$_{\text{obs}}$ 
versus the cumulative N$_{\text{pulse}}$ is plotted in panel (d). In this diagram, the slope represents the local values of 
detected pulse rate. This can also be seen in panel (c) of Figures \ref{tobs_npulse_1819}, \ref{tobs_npulse_1841} and \ref{tobs_npulse_1913}, 
in which we observe that the cumulative T$_{\text{obs}}$ and the 
cumulative N$_{\text{pulse}}$ does not always have same slope. The rate of pulses vary significantly between epochs. This is in agreement with the 
inference from Figure \ref{rate}. Table \ref{tab:burst_stat} lists the average pulse rates for the three RRATs for the full time span of observations 
and for smaller time spans, showing the evolution of average pulse rate with time. 
The average rate of pulses/hour (from Table \ref{tab:burst_stat}) are plotted in panel (e) of Figures \ref{tobs_npulse_1819}, \ref{tobs_npulse_1841} 
and \ref{tobs_npulse_1913}. The average rates are similar to the fitted detection rate from Table \ref{tab:fig1_burst_stat} as expected. These also 
indicate a marginal increase in the detection rate observed for PSR J1819$-$1458 and PSR J1913$+$1330, and roughly constant detection rates 
for PSR J1840$-$1419.  

\subsection{Timing of J1819$-$1458} \label{sec:timing_J1819}
\begin{center}
\begin{figure*}
\begin{center}
\includegraphics[angle=-90,width=1.2\textwidth]{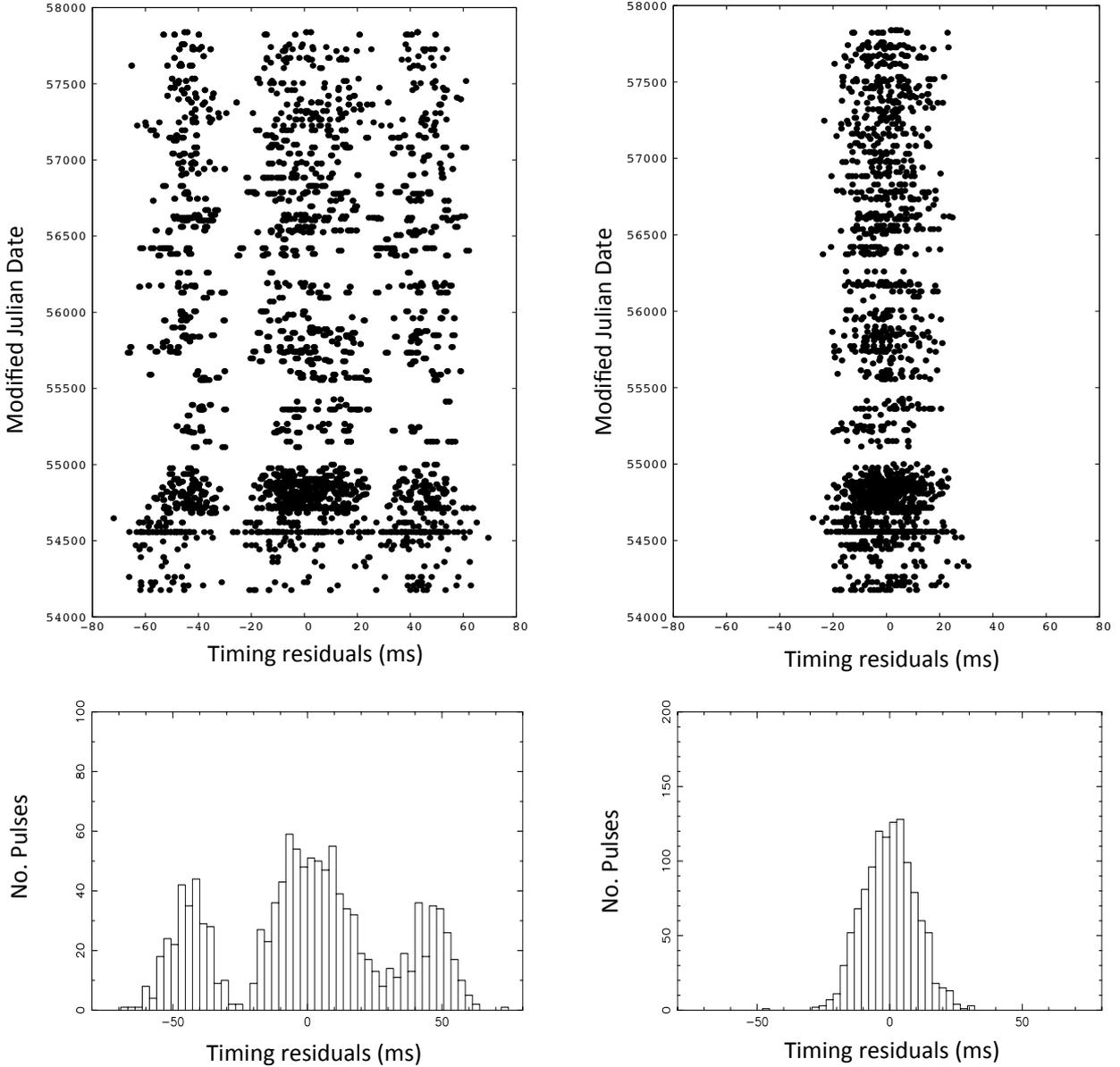}
\caption[Banded and unbanded residual]{Left panel: The timing residuals from the TOAs from the Lovell and the Parkes observations at L-band of the 
individual pulses from PSR J1819$-$1458 from post-glitch observations (relative to the model in Table \ref{tab:residuals_1819}) for 
$\sim$ 3900 days i.e. $\sim$ 11 years, showing that the majority of TOAs are located in three clearly identifiable bands with the accumulated 
arrival time histograms shown below. Right panel: The same residuals, but with the TOAs in the early and late bands fitted using offsets 
of $-$43.2$\pm$1.5 ms and $+$46.1$\pm$1.4 ms relative to the central band, to produce unbanded residuals, with the accumulated arrival 
time histogram shown below. The rms of the residuals subsequently decreases from 31.4 to 8.9 ms.}
\label{residual}
\vspace{0.1cm}
\end{center}
\end{figure*}
\end{center}

Figure \ref{residual} presents the timing residuals for all the pulses (above 5$\sigma$ detection limit) from PSR J1819$-$1458 from 
September 2006 till August 2017, i.e. for $\sim$ 11 years. 

We observe that the pulsed-emission from PSR J1819$-$1458 is grouped within three separated longitude regions covering about 120 ms of pulse phase, 
which is also clearly seen in a histogram of the residuals (bottom left panel of Figure \ref{residual}). The central band consists of 53$\pm$2\% 
of the detected pulses whereas the early and late bands consist of 26$\pm$1\% and 21$\pm$1\% of pulses respectively. The three band structure is 
consistent with the observations from \cite{lyne09}. To uniquely identify the three bands, we have considered data in three phase regions and 
separately fitted a model to identify the time offsets of three bands. To determine the band offsets, we put JUMP commands around 
TOAs in early and late bands, and fitted using tempo with few spin-frequency 
derivatives as required to whiten the timing data. This resulted in offsets of $-$43.2$\pm$1.5 ms between the central and the early band and 
$+$46.1$\pm$1.4 ms between the central and late band respectively. The measured offsets are consistent with the $\pm$ 45 ms offset used in \cite{lyne09}. 
The right panel of Figure \ref{residual} shows the residuals with the three bands aligned with these offsets. The rms of the 
residuals decreases from 31.4 ms for banded TOAs to 8.9 ms for the unbanded aligned TOAs. 
As a result of this procedure, the residuals are improved by a factor of 3.5 and uncertainties in the fitted parameters are 
similarly reduced. 

\begin{figure}
\includegraphics[angle=-90,width=1.0\textwidth]{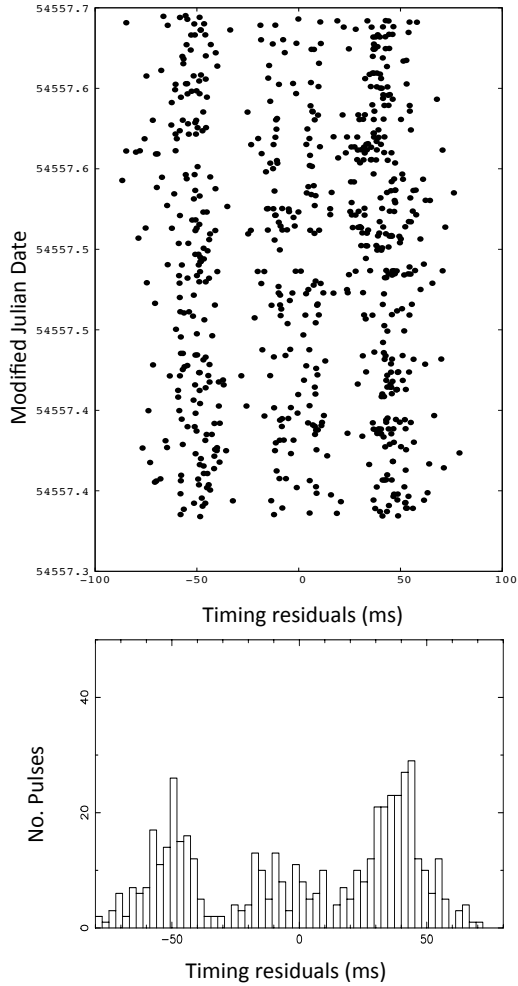}
\caption[GBT residuals]{The timing residuals from the TOAs of the individual pulses from PSR J1819$-$1458, from the GBT observations at 2.2 GHz on 1st April 2008, (relative to the model in Table \ref{tab:residuals_1819}) showing that TOAs at this frequency are also distributed in bands (top panel). The rms residual is 40.6 ms. The histogram of the residuals is shown in the bottom panel.}
\label{residual_gbt}
\vspace{0.1cm}
\end{figure}
\begin{figure*}
\begin{center}
\includegraphics[angle=-90,width=0.95\textwidth]{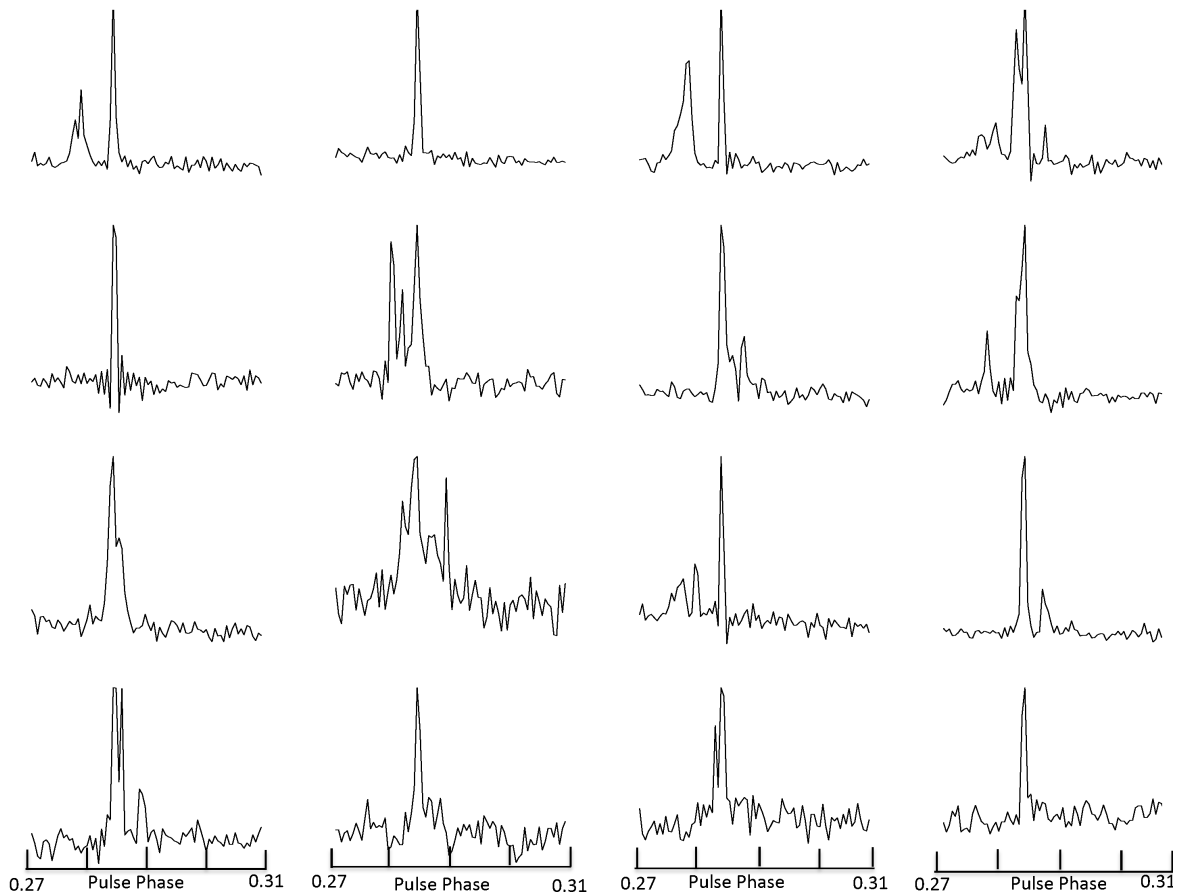}
\vspace{0.1cm}
\caption{A selection of single pulses detected from PSR J1819$-$1458, showing that individual pulses vary greatly and have complex profile structures.}
\label{sp_1819}
\vspace{0.1cm}
\end{center}
\end{figure*}
\begin{figure}
\begin{center}
\includegraphics[angle=0,width=0.5\textwidth]{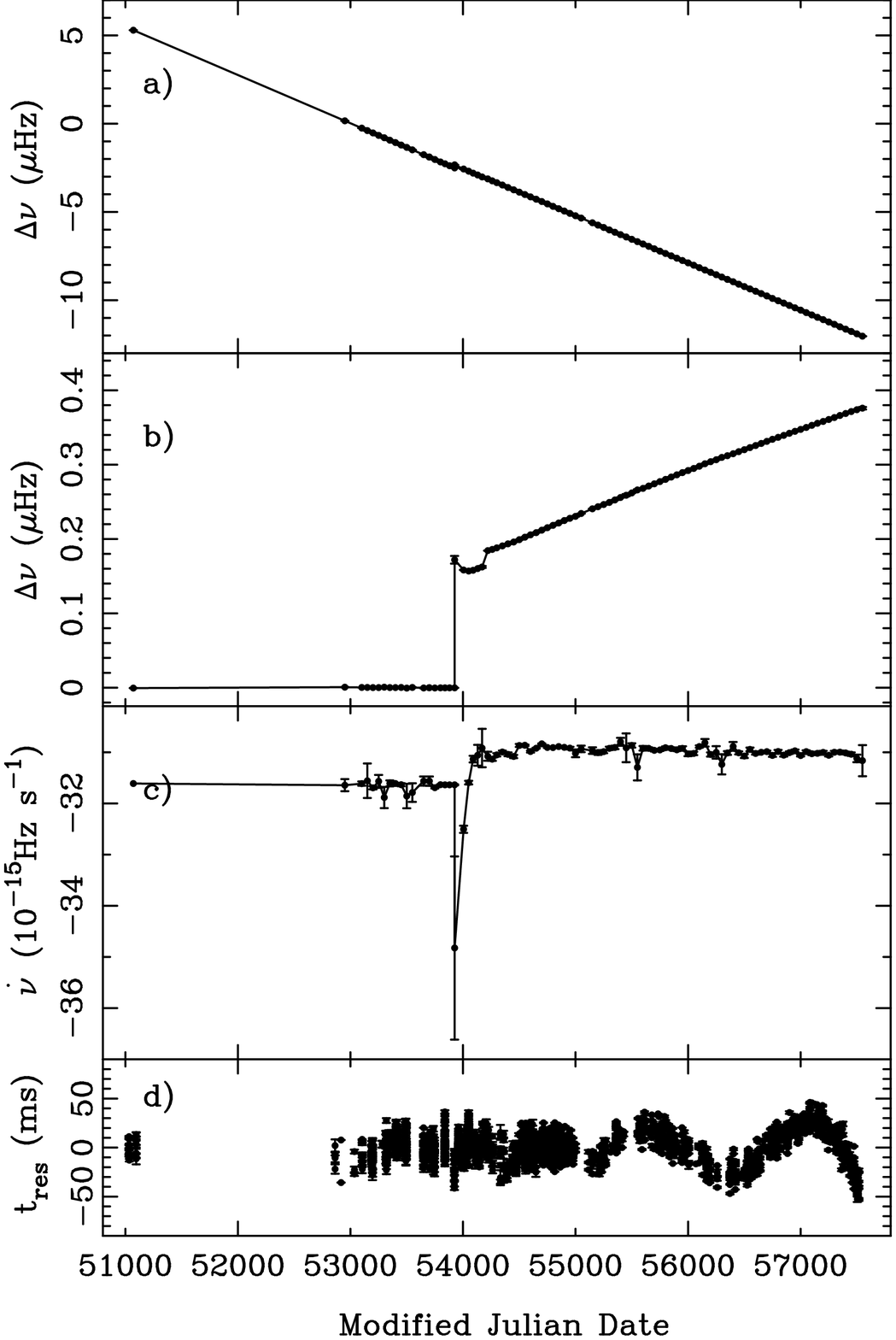}
\caption{Rotational evolution of PSR J1819$-$1458 over about an 18-year duration: (a) the spin-down in rotation rate of the RRAT, 
interrupted by a major glitch at MJD $\sim$53924, (b) the frequency residuals relative to a simple spin-down model fitted to 
data between 51000 and 53900, showing the second, relatively smaller, glitch at MJD $\sim$54168,
(c) the frequency derivative $\dot\nu$, showing the rapid increase in the magnitude of the spin-down rate immediately after the 
glitch, followed by a quasi-exponential decay and ultimately a gradual and slow increase of the slow down rate. 
(d) the timing residuals relative to the rotational model given in Table \ref{tab:residuals_1819}. The noisier residuals near 
the end of the data set are possibly due to a change in $|\dot\nu|$.}
\label{fig_1819_glitch}
\vspace{0.1cm}
\end{center}
\end{figure}
Figure \ref{residual_gbt} shows the TOAs of PSR J1819$-$1458 from GBT observations at 2.2 GHz on 1st April 2008, with an rms of the 
residuals of 40 ms. The separation between the two outer bands is 98 $\pm$ 4 ms, which is 9$\pm$1\% more than the 1.4 GHz separation. This is 
indicative of a wider emission region at 2.2 GHz than 1.4 GHz, which is in the opposite direction to that predicted by radius to 
frequency mapping \citep{cordes78}. In addition we observe that at 2.2 GHz, unlike 1.4 GHz, the majority of the pulses are not 
from the central band and the central band TOAs are frequently split in two bands. 
Figure \ref{sp_1819} shows examples of single pulses from PSR J1819$-$1458, with a few having complex profiles including single,
double and triple  peaks.
  
\begin{table}
\centering
\caption{Timing parameters of PSR J1819$-$1458
\label{tab:residuals_1819}}
\begin{tabular}{|l|c||c|c|}
\hline
Pre-glitch parameters \citep{lyne09}& \\
  \hline
Right ascension (J2000)\dotfill &18$^\mathrm{h}$19$^\mathrm{m}$34\fs173$\dagger$                \\
Declination (J2000)\dotfill     &$-$14\degr58\arcmin03\farcs57$\dagger$            \\
Pulsar frequency $\nu$(s$^{-1}$)\dotfill  &0.23456756350(2)                               \\
Pulsar frequency derivative $\dot\nu$ (s$^{-2}$)\dotfill &$-$31.647(2)$\times$10$^{-15}$  \\
Period epoch (MJD)\dotfill                                & 54451                                   \\
Timing data span (MJD)\dotfill                            & 51031$-$54938                         \\
Dispersion measure $\mbox{DM}$ (pc~cm$^{-3}$)\dotfill      & 196.5                                 \\
Post-fit residual rms (ms)\dotfill                         & 10.2                           \\\hline
Glitch 1 parameters & \\\hline
Epoch (MJD)\dotfill                                       & 53924.79(15) \\
Incremental $\Delta \nu$ (Hz)\dotfill                     & 0.1380(6)$\times$10$^{-15}$\\\hline
Glitch 2 parameters & \\\hline
Epoch (MJD)\dotfill                                       & 54168.6(8) \\
Incremental $\Delta \nu$ (Hz)\dotfill                     & 0.0226(3)$\times$10$^{-6}$\\\hline
Post-glitch parameters \\\hline
Right ascension (J2000)\dotfill &18$^\mathrm{h}$19$^\mathrm{m}$34\fs16(1)                \\
Declination (J2000)\dotfill     &$-$14\degr58\arcmin00\farcs00(1)            \\
Pulsar frequency $\nu$(s$^{-1}$)\dotfill                          & 0.234564843(4)                               \\
Pulsar frequency derivative $\dot \nu$ (s$^{-2}$)\dotfill         & $-$30.959(4)$\times$10$^{-15}$  \\
Pulsar frequency second derivative $\ddot\nu$ (s$^{-3}$)\dotfill  & $-$1.24(2)$\times$10$^{-24}$       \\
Pulsar frequency third derivative $\dddot \nu$ (s$^{-4}$)\dotfill & $-$2.4(6)$\times$10$^{-33}$       \\
Pulsar frequency forth derivative $\ddddot\nu$ (s$^{-5}$)\dotfill & $-$3.0(8)$\times$10$^{-39}$       \\
Pulsar frequency fifth derivative  (s$^{-6}$)\dotfill & $-$1.1(4)$\times$10$^{-47}$       \\
Period epoch (MJD)\dotfill                                & 55996.24                                    \\
Timing data span (MJD)\dotfill                            & 54175.87$-$57838.37                         \\
Dispersion measure $\mbox{DM}$ (pc~cm$^{-3}$)\dotfill      & 196.5                                 \\
Number of TOAs\dotfill                                    & 1373                          \\
Post-fit residual rms (ms)\dotfill                         & 8.9                           \\\hline

Derived parameters &\\
  \hline
Period (s) \dotfill    & 4.2632901504(1)\\
Period Derivative \dotfill & 5.62717(4)$\times$10$^{-13}$\\
Braking Index from $\nu$,$\dot\nu$,$\ddot\nu$ \dotfill & $-$226 \\
Total time span (yr) \dotfill & 10.03 \\
Spin down energy loss rate $\dot{E}$ (erg/s)\dotfill       &2.8$\times$10$^{32}$                  \\
Spin down age (yr)\dotfill                                 &1.2$\times$10$^{5}$             \\
Surface magnetic flux density (Gauss)\dotfill              &4.9$\times$10$^{13}$           \\
DM distance$^\ddagger$ (kpc)\dotfill                        & 3.3                   \\\hline
\end{tabular}
\\
$\dagger$ {errors are 2$\arcsec$ as derived from Chandra observations \citep{rea09}}\\
$\ddagger$ using \cite{yao17} model of electron distribution\\
\end{table}

\subsubsection{Post-glitch frequency evolution}  \label{sec:glitch_J1819}
Glitches are sudden jumps of rotational period and are detected as a result of regular monitoring of a pulsar. A timing model fitting 
$\nu$, $\dot\nu$, $\ddot\nu$ to the pre-glitch TOAs usually describes the pulsar rotation, but after a glitch one needs to have a new 
timing model for fitting the post-glitch TOAs. In the timing campaign described in \cite{lyne09}, they detected two glitches 
at MJD $\sim$53924 and $\sim$54168, and studied the post-glitch timing properties for $\sim$800 days after the glitches. 
In the present work, we find no further glitches, but carry out further investigation of the post-glitch rotational properties of 
PSR J1819$-$1458 for about 3700 days.  
Table \ref{tab:residuals_1819} presents the pre-glitch and post-glitch timing models for PSR J1819$-$1458. Figure \ref{fig_1819_glitch} 
shows the frequency evolution of PSR J1819$-$1458 over about 18 years. Panel (a) shows the slow down of the RRAT and the large glitch 
observed at MJD $\sim$53924. We fit a simple slow down model (fitting only pulsar frequency and its derivative) to the data between 
MJD 51000 and 53900 and a second relatively smaller glitch 
is now visible in panel (b) at MJD $\sim$54168. The post-glitch time evolution of the frequency derivative $\dot \nu$ plotted in panel (c), can be 
classified in a few stages: \\
(i) Rapid increase of $|\dot \nu|$: a rapid increase of $|\dot \nu|$ is observed immediately after the glitch, $|\dot \nu|$ 
is $\sim$31.6 $\times$ 10$^{-15}$ Hz s$^{-1}$ before the glitch and immediately after the glitch $\dot \nu$ increases 
to $\sim$35.0$\times$ 10$^{-15}$ Hz s$^{-1}$. \\ 
(ii) Post-glitch recovery of $|\dot \nu|$: Next $|\dot \nu|$ exponentially decreases.\\
(iii) Over-recovery of the $|\dot \nu|$: $|\dot \nu|$ reaches an asymptotic value 
of $\sim$ 31.0 $\times$ 10$^{-15}$ Hz s$^{-1}$, which is significantly smaller than the pre-glitch value 
of $\sim$31.6 $\times$ 10$^{-15}$ Hz s$^{-1}$.\\
(iv) Recovery from over-recovery of $|\dot \nu|$: After MJD $\sim$55000, the $|\dot \nu|$ again starts to increase and 
reaches $\sim$31.1 $\times$ 10$^{-15}$ Hz s$^{-1}$ at the time of writing this paper.\\
Glitches observed in other pulsars are also charatersised by stages like (i) and (ii). However, for PSR J1819$-$1458 we observe
that the recovery of the frequency derivative goes beyond the pre-glitch value and we observe stages (iii) and (iv), which is not the case 
for the other pulsars (aside from PSR J1119$-$6127, \cite{weltevrede11}).
The implication of this result and comparison 
with the post-glitch timing properties of normal pulsars are discussed further in \S \ref{sec:discussion}.

\begin{figure*}
\includegraphics[angle=-90,width=1.2\textwidth]{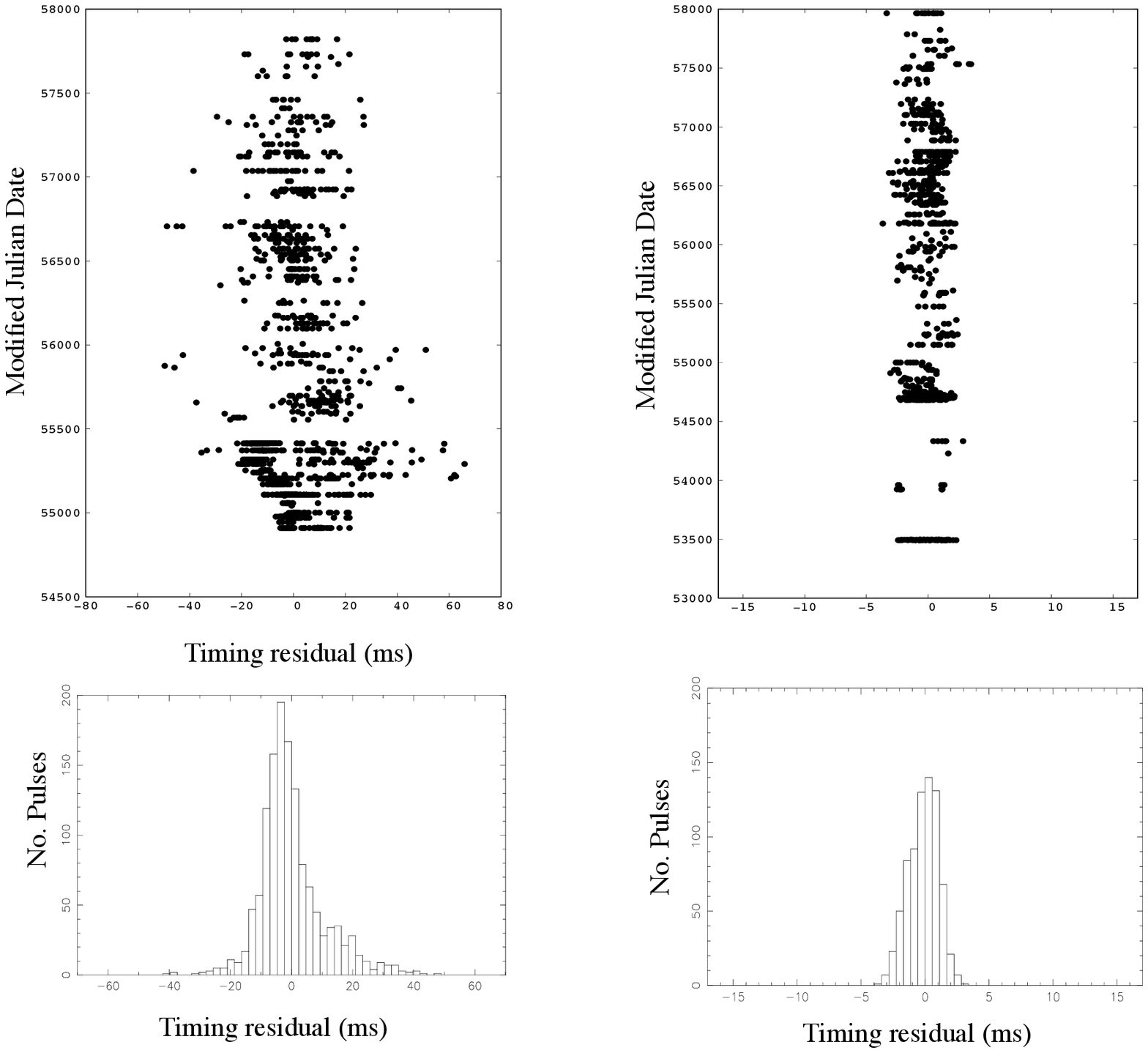}
\caption[J1840 residual]{Left panel: The timing residuals from the TOAs of the individual pulses from PSR J1840$-$1419 for $\sim$2900 days 
i.e. $\sim$ 8 years, relative to the timing model given in Table \ref{tab:residuals}, fitting for pulsar frequency $\nu$, its first two 
derivatives and pulsar position, resulting in residuals with an rms value of 12.6 ms. The histogram of the corresponding timing residuals is 
shown in the bottom panel. Right panel: The timing residuals from the TOAs of the individual pulses from PSR J1913$+$1330 for $\sim$ 4480 days 
i.e. $\sim$ 12 years, relative to a slow down model given in Table \ref{tab:residuals}, fitting pulsar frequency $\nu$, its first five frequency 
derivatives and pulsar position, giving an rms of 1 ms. The histogram of the corresponding timing residuals is shown in the bottom panel.}
\label{residual_J1913_J1841}
\vspace{0.1cm}
\end{figure*}

\begin{table*}
\centering
\caption{Timing parameters of PSR J1840$-$1419 and J1913$+$1330
\label{tab:residuals}}
\begin{tabular}{|l|c||c|}
\hline
Parameters  & J1840$-$1419 & J1913$+$1330\\
  \hline
Right ascension (J2000)\dotfill    &18$^\mathrm{h}$40$^\mathrm{m}$33\fs04(1)         &19$^\mathrm{h}$13$^\mathrm{m}$17\fs97(1)        \\
Declination (J2000)\dotfill        &$-$14\degr19\arcmin06\farcs5(9)                  &$-$13\degr30\arcmin32\farcs78(4)               \\
Pulsar frequency $\nu$(s$^{-1}$)  &0.151571128974(2)                                &1.0829644010729(3)  \\
Pulsar frequency derivative $\dot\nu$ (s$^{-2}$)\dotfill &$-$1.4597(3)$\times$10$^{-16}$  &$-$1.01772(2)$\times$10$^{-14}$ \\
Pulsar frequency double derivative $\ddot\nu$ (s$^{-3}$)\dotfill   &$-$1.6(1.4)$\times$10$^{-27}$     &6(5)$\times$10$^{-27}$                 \\
Pulsar frequency triple derivative $\dddot\nu$ (s$^{-4}$)\dotfill  &  $-$                       &$-$6.9(3)$\times$10$^{-33}$                  \\
Pulsar frequency forth derivative $\ddddot\nu$ (s$^{-5}$) \dotfill  &  $-$                       &7(1)$\times$10$^{-41}$                 \\
Period epoch (MJD)\dotfill                                   & 55074.9                    & 55090.9            \\
Timing data span (MJD)\dotfill                               & 54909.889$-$57820.378      & 53491.80$-$57964.82                \\
Dispersion measure $\mbox{DM}$ (pc~cm$^{-3}$)\dotfill        & 20.0                       & 175.6          \\
Number of TOAs\dotfill                                       & 1438                       & 815\\\hline
Post-fit residual rms (ms)\dotfill                           & 12.6                       & 1.1  \\\hline
  \multicolumn{3}{c}{Derived parameters} \\
  \hline
Period (s) \dotfill                                          & 6.5975625223(1)             & 0.923391386650(2)                    \\
Period Derivative \dotfill                                   & 6.353(1)$\times$10$^{-15}$  & 8.6776(2)$\times$10$^{-15}$                  \\
Braking Index from $\nu$,$\dot\nu$,$\ddot\nu$ \dotfill                  & $-$11986                    & 63.54                  \\
Total time span (yr) \dotfill                                & 7.97                        & 12.46                  \\
Spin down energy loss rate $\dot{E}$ (erg/s)\dotfill         & 8.7$\times$10$^{29}$        &4.2$\times$10$^{32}$       \\
Spin down age (yr)\dotfill                                   & 1.6$\times$10$^{7}$         &1.6$\times$10$^{6}$ \\
Surface magnetic flux density (Gauss)\dotfill                & 6.5$\times$10$^{13}$        &2.8$\times$10$^{12}$ \\
DM distance$^\dagger$ (kpc)\dotfill                          & 0.73                  & 6.1\\\hline
\end{tabular}
\\
$\dagger$ using \cite{yao17} model of electron distribution.
\end{table*}

\subsection{Timing of PSR J1840$-$1419} \label{sec:timing_J1840}
The left panel of Figure \ref{residual_J1913_J1841} shows the timing residuals from TOAs of the individual pulses from PSR J1840$-$1419 over 
$\sim$ 8 years relative to the slow down model in Table \ref{tab:residuals}. This model obtained by fitting for $\nu$, $\dot \nu$, 
$\ddot \nu$ and pulsar position results in timing residuals with an rms of 12.6 ms. Although the observed spread in the residuals 
is $\sim$ $\pm$ 40 ms, the majority of TOAs are within $\sim$ $\pm$ 20 ms. 
The histogram of the timing residuals is shown in the bottom panel.

\subsection{Timing of PSR J1913$+$1330} \label{sec:timing_J1913}
\begin{figure*}
\begin{center}
\includegraphics[angle=-90,width=1.0\textwidth]{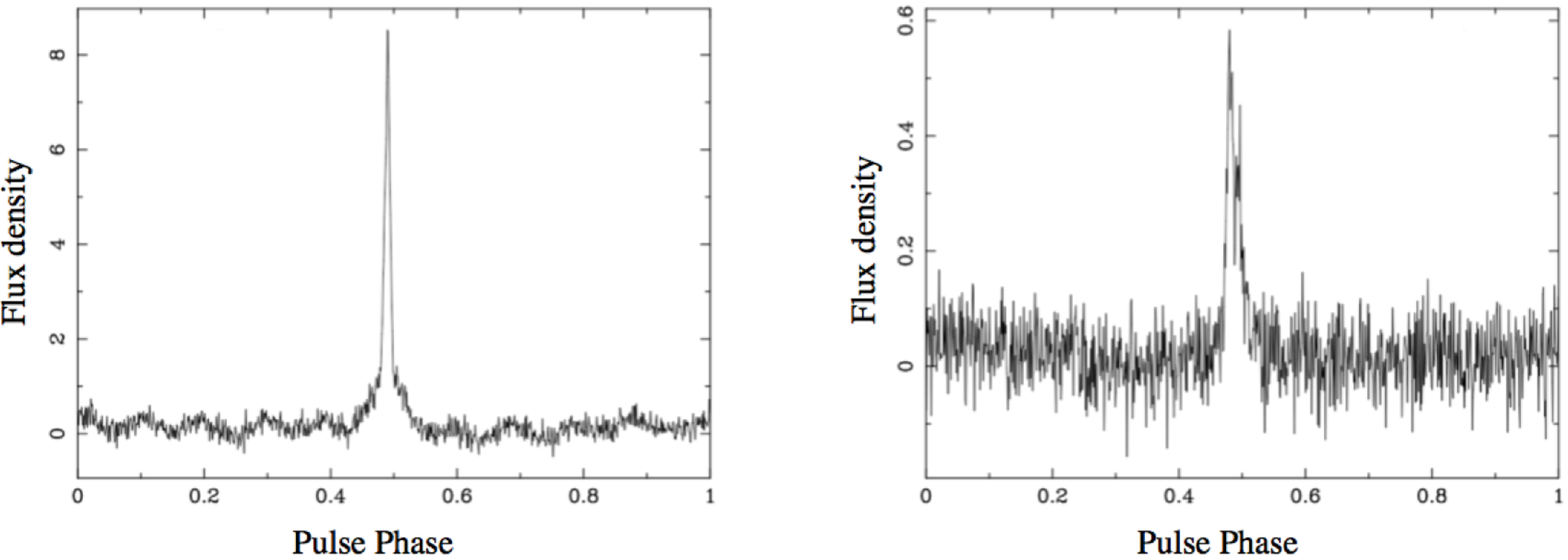}
\vspace{-2cm}
\caption[rrat and weak pulses J1913$+$1330]{Left panel: Average profile of the RRAT pulses detected for PSR J1913$+$1330 ($\sim$540 pulses). Right panel: Average profile of the weak mode detected for the PSR J1913$+$1330 ($\sim$7000 pulses). Flux density is in arbitrary units, but the two plots are on the same scale.}
\label{fig:avp_rrat}
\end{center}
\end{figure*}

The right panel of Figure \ref{residual_J1913_J1841} shows the timing residuals of PSR J1913$+$1330 over $\sim$12 years, relative 
to a timing model given in Table \ref{tab:residuals} derived after fitting for $\nu$ and its first four derivatives and pulsar 
position, resulting in timing residuals with an rms of 1.0 ms. The histogram of the timing residuals is shown in the bottom panel. 
The right panel of Figure \ref{fig:avp_rrat} presents the average profile of the RRAT pulses detected for J1913$+$1330 (created with 
the {\sc psrsalsa} software package \cite{weltevrede16}). 
\subsection{Weak emission mode for PSR J1913$+$1330} \label{sec:average_det_J1913}
\begin{figure}
\begin{center}
\includegraphics[angle=0,width=0.5\textwidth]{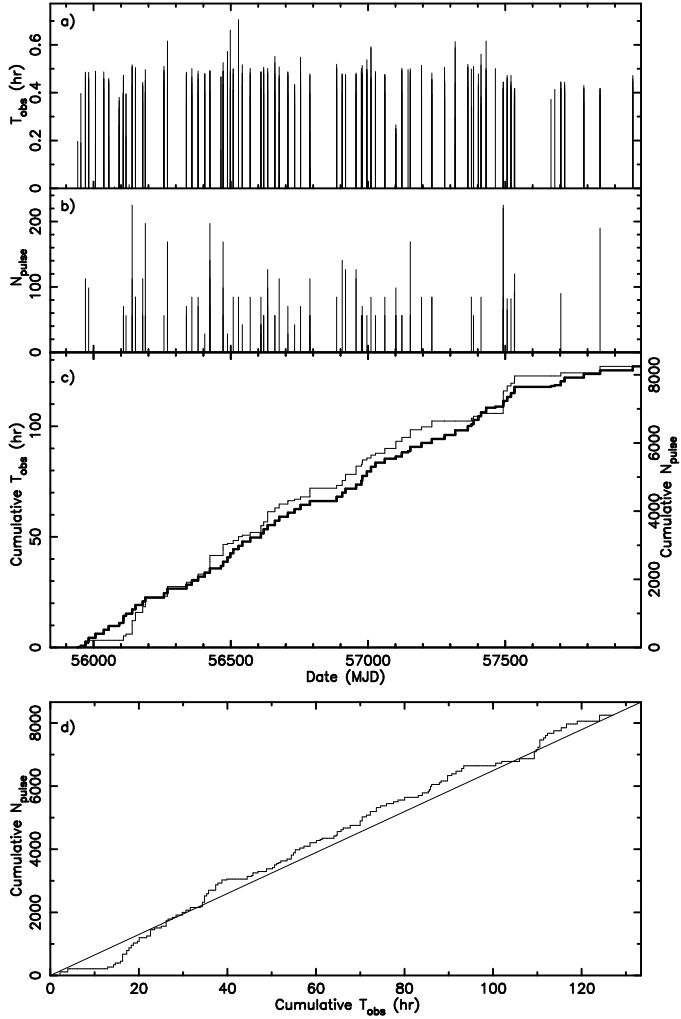}
\caption[tobservation and number of pulses J1913$+$1330]{Average emission statistics of PSR J1913$+$1330 for the weak mode. (a) Duration of observation (T$_{\text{obs}}$) vs the epoch of observation (in MJD) in which average emission from the RRAT was observed, (b) Number of pulses (N$_{\text{pulse}}$) equivalent of average emission detected vs the epoch of observation (MJD), (c) Cumulative T$_{\text{obs}}$ (with heavy line)
and Cumulative N$_{\text{pulse}}$ (with thin line) vs the epoch of observation, (d) Cumulative T$_{\text{obs}}$ vs Cumulative N$_{\text{pulse}}$, solid line represents
the mean emission rate considering the start and stop range of observations.}
\label{tobs_npulse_1913_roach}
\vspace{0.1cm}
\end{center}
\end{figure}

In addition to the RRAT pulses, we report the detection of a persistent but weak emission mode for PSR J1913$+$1330.
The weak mode is observed after averaging pulses together and is followed by a long absence of any detectable emission.
We marked time slices with a pulse detected at more than 5$\sigma$ signal-to-noise in a 1 minute integration as a detection 
of the weak mode emission. The duration of the weak emission mode varied from 2 minutes to 14 minutes during our observations. 
Interestingly, strong RRAT pulses were not present during the weak mode intervals. The left panel 
of Figure \ref{fig:avp_rrat} presents the average profile of the pulses detected in its weak mode. 
Though the average profile for the burst mode is single-peaked, we see a double peaked-profile for the weak emission mode. 
The mean flux density of the pulses in the weak emission mode is lower than the mean of the RRAT pulses by about a factor of 50.
Figure \ref{tobs_npulse_1913_roach} plots the emission statistics of this weak average emission mode which can be 
compared to the emission statistics of bright single pulses typically seen for the RRATs (Figure \ref{tobs_npulse_1913}).
Table \ref{tab:burst_stat} compares the rate of emission (in pulses/hr) for the RRAT mode and the weak emission mode. The average 
rate of pulse emission in the weak mode is $\sim$64 pulses/hr, which is at least an order of magnitude higher than the rate of 
emission in RRAT mode. This assumes that all pulses accumulated in the weak mode have similar strengths. With this assumption, the 
total detection of pulses in the weak mode translates to $\sim$7000 pulses in $\sim$110 hours of observation. This indicates that 
the weak mode emission is detected for 1.6\% of the total observing duration. We note that PSR J1913$+$1330 emits bright single 
pulses typical for RRATs for 0.1\% of the total observing duration. 
\section{Discussion and summary}\label{sec:discussion}
We now discuss the main outcomes and the corresponding implications of this work.\\
\subsection {Pulse rate statistics} 
We have studied pulse rate statistics using the data from the Lovell telescope for three RRATs.  
We report a possible long-term increase in the emission rate for PSR J1819$-$1458 (Figure \ref{rate}). We also see evidence for a marginal 
increase in pulse emission rate for J1913$+$1330, but for PSR J1840$-$1419 no long-term change in emission rate is observed. For PSR J1840$-$1419, 
\cite{keane11} determined a RRAT pulse emission rate of $\sim$60 pulse/hour for a study during MJD 54909$-$55239 using the Parkes telescope, 
which is considerably higher than the average of $\sim$25 pulse/hour from our study with the Lovell telescope possibly due to the use of a 
smaller bandwidth and a worse RFI environment. For PSR J1913$+$1330, 
\cite{mclaughlin09} determined a rate of $\sim$1.5 pulses/hour during MJD 53035$-$54938 with the Lovell telescope. This differs by at least 
a factor of two from the average detection rate of $\sim$4.9 pulses/hour for the duration of MJD 55150$-$57400 listed 
in Table \ref{tab:burst_stat}. This may be due to use of the more sensitive DFB backend during our study compared to the narrower band AFB, or 
may indicate a long-term increase in the pulse rate. 

We report significant variations of the pulse rates between the observing epochs. 
We reported emission rates of $\sim$230 pulses/hour and $\sim$123 pulses/hour for two observing epochs for PSR J1913$+$1330. However,
since both these epochs are relatively short ($\sim$6 mins) compared to the typical 30 mins observing epochs, it is likely that we managed to 
hit on a period of time when the RRAT was active. 
Variations of two orders of magnitude from the mean pulse rates at individual epochs has implications for possible RRAT emission models. 
Giant pulses from weak pulsars is one of the possible explanations for RRAT emission. \cite{lundgren95} reported that though the observed 
rate of giant pulse emission from the Crab pulsar changes from day to day, above a fixed threshold the rate of all the giant pulses emitted 
remains fixed and observed variability is caused by propagation effects in the interstellar medium. Therefore, whether the observed two orders of 
magnitude variation in pulse detection rate for PSR J1913$+$1330 can be explained by propagation effects may influence the feasibility 
of giant pulse like origin of the RRAT pulses. 
The other proposition of for RRAT behaviour being extreme nulling of radio pulsars a with randomly varying active and null state is still 
a feasible mechanism. 
Circumstellar asteroid belts around the pulsar \citep{cordes08} could be feasible subject to the requirement to explain such 
highly varying emission rates.

\subsection{Long-term timing of RRATs} 
We present long-term radio timing results for RRATs J1819$-$1458, J1840$-$1419 and J1913$+$1330.
For timing of most pulsars we use integrated profiles, which are stable and phase stability is implicitly assumed.
However, for single pulse timing this assumption is not valid and extra scatter in the timing residuals is expected and 
is clearly seen in Figure \ref{residual} and \ref{residual_J1913_J1841}. 
For PSR J1819$-$1458 we have seen that the shape of the individual pulses varies greatly, which is also commonly seen for normal pulsars.
However, the distribution of pulse arrival times from PSR J1819$-$1458 at 1.4 and 2.2 GHz indicate a trend opposite to the radius to frequency 
mapping generally followed by pulsars. It will be intriguing to study the frequency-evolution of the separation of profile components over a wider 
frequency and with simultaneous data. 

This is the longest time-span study performed for RRATs so far and therefore enables us to compare the long-term timing properties 
of the RRATs with the other pulsars. Figure \ref{ppdot_rrats} shows the RRATs studied in this paper in the $P-\dot P$ diagram along with 
other pulsars and RRATs. We find that the long-term timing properties of these RRATs are similar to the other pulsars. We note that the 
estimated surface magnetic field strength of J1819$-$1458 is the highest known among RRATs. 

For the pulsars that emit giant pulses, the inferred magnetic field strength at the light cylinder 
$B_{\text{LC}}\propto$ $P^{−2.5}~\dot{P}^{0.5}$ is an 
indicator of giant pulse emissivity (e.g \cite{knight06}, \cite{cognard04}), with $B_{\text{LC}}$ $>$ 10$^5$ G for the giant pulse 
emitting pulsars. However, \cite{knight06} argue that $\dot{E}\propto$ $P^{-3}~\dot{P}$ rather than $B_{\text{LC}}$ may be a better 
indicator of giant pulse emission. $B_{\text{LC}}$ and $\dot{E}$ of the three RRATs studied in this paper (inferred from 
Tables \ref{tab:residuals_1819} and \ref{tab:residuals}) are many order of magnitude less than the proposed values. 
Moreover, as the pulses detected from RRATs are much broader ($\sim$ milliseconds) than the traditional giant pulses ($\sim$ nano-seconds), 
this argues against RRAT pulses being a manifestation of giant pulse emission. \\

\subsection{Post-glitch timing properties of PSR J1819$-$1458} 
We studied the timing properties of J1819$-$1458 for over 6500 days, and report unique post-glitch timing 
properties for about 3700 days after the glitch at MJD 54167. 
A long-term decrease of $|\dot \nu|$ following the glitch is observed, implying that the pulsar position 
in the $P-\dot P$ diagram shifts vertically downwards after the glitches 
as reported by \cite{lyne09}. Glitches observed for other pulsars result in an abrupt increase 
of $|\dot \nu|$ during the glitch, which then decreases after the glitch and stabilises resulting in a 
long-term increase in spin-down rate. For example, \cite{espinoza11} presented a database of 315 glitches 
from 102 pulsars and showed that the result of a frequency-glitch in normal pulsars is a net increase in 
slow down rate (Figure 6 of \cite{espinoza11}) and an upwards step in the $P-\dot P$ diagram. Repeated 
occurrence of such glitches involving a long-term decrease of $|\dot \nu|$, in J1819$-$1458, would imply 
that this RRAT will gradually move from magnetar-like spin properties to those of radio pulsars. Since it is difficult to explain 
the observed post-glitch evolution of $\dot \nu$ with the conventional model of sudden unpinning of the vortex lines and subsequent 
transfer of angular momentum from the super-fluid to the crust, \cite{lyne09} pointed out that observed glitches 
in PSR J1819$-$1458 could be magnetar like. 
Such glitches are frequently observed for the magnetars, and are thought to originate due to the high internal magnetic field 
that can deform or crack the crust \citep{thompson96}.

PSR J1119$-$6127 is another radio pulsar to show a similar post-glitch long-term decrease of $|\dot \nu|$ \citep{weltevrede11}. 
Incidentally PSR J1119$-$6127 also has a high surface magnetic field (B$\sim$4.1$\times$10$^{13}$ G), like 
PSR J1819$-$1458 (B$\sim$4.94$\times$10$^{13}$ G). \cite{antonopoulou14} termed such a peculiar post-glitch behaviour 
as ``over-recovery'' of the spin-down rate and suggested that they were magnetar-like glitches. Recently, \cite{archibald16} have reported 
a magnetar like outburst from this pulsar. Similar ``over-recovery'' in frequency is also reported for a X$-$ray pulsar 
J1846$-$0258 \citep{livingstone2k10}. It also has a relatively high inferred magnetic field (B$\sim$5$\times$10$^{13}$ G). Such 
interesting magnetar like properties of high magnetic field pulsars, and similarity in glitch properties with PSR J1119$-$6127, 
emphasise the importance of regular monitoring of PSR J1819$-$1458. 

After the episode of ``over-recovery'' immediately after the glitch for PSR J1819$-$1458, we observe a  
very slow ``recovery from the over-recovery'' (i.e. $|\dot \nu|$ again starts to increase consistently) starting significantly 
later ($\sim$ 1000 days after the glitch episode) and continuing untill the point of writing this paper.  It is possible that 
eventually the pre-glitch $\dot \nu$ value will be reached with such a recovery process if it is not interrupted by another glitch. 
\cite{lyne09} commented that for PSR J1819$-$1458 the 
spin-down rate will decay to zero on a time scale of few thousand years if the pulsar underwent similar glitches every 30 years 
resulting in a permanent decrease in slow-down rate (i.e. a step down in the $P-\dot P$ diagram).
However, the ``recovery from the over-recovery'' observed by us for this pulsar will play a major role in deciding how the slow down rate 
will evolve and the predicted time for the spin-down rate to reach to zero, if at all. 
It is also possible that before the next glitch the ``recovery from the over-recovery'' places the RRAT at its original position 
of the $P-\dot P$ diagram. We note that even $\sim$3700 days after the glitch, the effects of the glitch persist, indicating 
a very long-term memory of the process. Theoretical models explaining the glitch phenomena will be constrained by this and it will be 
interesting to see if the occurrence of the next glitch is random or has some relation with the recovery process. 
Moreover, since PSR J1819$-$1458 is the only RRAT for which glitches are observed, it will be interesting to know if these glitches 
are representative of RRATs. This can only be verified with regular monitoring to detect possible glitches in other RRATs. 

\subsection{Weak emission mode for PSR J1913$+$1330}
In addition to regular active and off modes observed for RRATs, we have detected a second weak emission mode for PSR J1913$+$1330
(detailed in \S \ref{sec:average_det_J1913}), characterised by weak average emission followed by long absence of detectable emission. This 
is reminiscent of profile mode changing and nulling which are commonly observed for many pulsars. 
But PSR J1913$+$1330 is the only RRAT for which such weak emission is observed besides the normal active and off modes for RRATs. We also 
observe a difference in profile shape in the two modes. For the RRAT mode the average profile is single-peaked, whereas for the weak mode 
the profile is double-peaked. We find that the mean flux density of the RRAT pulses is $\sim$50 times 
higher than that of typical pulses in the weak emission mode. We also report that the total duration of the weak mode emission is at least 
an order magnitude higher than the total duration of RRAT single pulses. Finding a different mode of emission, similar to emission 
from normal pulsars, in RRATs has implications in understanding the connection between their emission processes. This indicates that 
the RRATs may be a manifestation of extreme nulling pulsars, in this case also with a very weak emission mode. 

The long-term study by \cite{young15} found that PSR J1853+0505 exhibits a weak emission state, in addition to its strong and null states. 
This indicates that nulls may represent transitions to weaker emission states which are below the sensitivity thresholds 
of particular observing systems. However, for most pulsars nulling is observed to be followed by emission for more than one pulse period. 
This is in contrary to the fact that most observed RRAT pulses are single. So RRATs could be a special manifestation of 
nulling that is not generally observed for normal pulsars.

In a study of PSR B0656$+$14, which was argued to have similar emission properties as RRATs if it is placed at a large distance, \cite{weltevrede06}, had postulated that longer observations 
of RRATs may reveal weaker emission modes in addition to the detected RRAT pulses. Detection of a weak emission mode for PSR J1913$+$1330 
may strengthen this hypothesis. However, it is noteworthy that, although the RRAT population have typically larger 
period and magnetic field strengths than the normal radio pulsar population (Figure \ref{ppdot_rrats}), the spin-down 
properties of PSR J1913$+$1330 are similar to those of the normal radio pulsar population. Thus PSR J1913$+$1330 is a special RRAT 
sharing properties of both the populations. Another similar pulsar is PSR J0941$-$39 which exhibits 
an RRAT-like emission rate of $\sim$90/100 pulses/hour at times and behaves like a strong pulsar with nulling at other times \citep{burke-spolaor10}.
Searching for such weak emission modes in other RRATs will be important in this context.

To conclude, we present the longest time span study of three RRATs. In addition, we described the detection-rate evolution, unusual post-glitch 
properties of PSR J1819$-$1458 and detected a pulsar-like emission mode for PSR J1913$+$1330. Instead of being a separate class of neutron star, RRATs 
can be manifestations of extreme emission types that are previously not seen in the rest of the neutron star population. Unraveling this will 
require comparison of timing properties of a large number of RRATs. But a large fraction of RRATs are not well studied, for example about 70\% 
of the known RRATs do not have a timing solution. 
Detailed long-term study of RRATs is warranted to establish the connection of RRATs with the rest of the neutron star population.
  
\section{Acknowledgments}
We thank reviewer of this paper for comments that helped us to improve the paper. 
B. Bhattacharyya acknowledges support of Marie Curie grant PIIF-GA-2013-626533 of European Union.
Pulsar research at Jodrell Bank centre for Astrophysics and access to the Lovell telescope is 
supported by a Consolidated Grant from the UK's Science and Technology Facilities Council. The Parkes 
radio telescope is part of the Australia Telescope National Facility which is funded by the Australian 
Government for operation as a National Facility managed by CSIRO. The Green Bank Observatory is a 
facility of the National Science Foundation operated under cooperative 
agreement by Associated Universities, Inc. We thank A. Holloway and R. Dickson of University of 
Manchester for making the Hydrus computing cluster at University of Manchester available for 
the analysis presented in this paper. \\ 






\appendix
\section{}
\label{sec:abbreviations}

\begin{table}
\caption{Variation in average burst rate for PSR J1819$-$1458, J1840$-$1419 and J1913$+$1330, for the epochs for which at least one pulse is detected.}
\label{tab:burst_stat}
\begin{tabular}{|l|c|c|c|c|c|c|c|}
\hline
  RRAT name    & MJD range          & Burst rate    \\
               &                    & (pulses/hr) \\\hline
  J1819$-$1458 & 55047$-$57436$^\dagger$  &15.5$\pm$0.5  \\
               & 55100$-$55300      &9.4$\pm$1.4  \\
               & 55300$-$55500      &6.2$\pm$1.0  \\
               & 55500$-$55700      &12.9$\pm$1.8  \\
               & 55700$-$55900      &14.2$\pm$1.5   \\
               & 55900$-$56100      &13.4$\pm$2.1 \\
               & 56100$-$56300      &22.9$\pm$2.9 \\
               & 56300$-$56500      &21.7$\pm$3.2 \\
               & 56500$-$56700      &19.2$\pm$1.9  \\
               & 56700$-$56900      &27.5$\pm$3.2 \\
               & 56900$-$57100      &16.2$\pm$2.0 \\
               & 57100$-$57300      &16.1$\pm$2.0 \\
               & 57300$-$57500      &16.3$\pm$1.5 \\
               & 57500$-$57700      &18.3$\pm$2.0 \\
               & 57700$-$57900      &16.5$\pm$1.4 \\
               & 57900$-$58000      &16.5$\pm$2.0 \\
\hline
  J1840$-$1419 & 55080$-$57377$^\dagger$   &24.1$\pm$0.7 \\
               & 55000$-$55200      &16.1$\pm$1.9 \\
               & 55200$-$55400      &28.3$\pm$2.6 \\
               & 55400$-$55600      &24.3$\pm$3.6 \\
               & 55600$-$55800      &29.7$\pm$2.5 \\
               & 55800$-$56000      &26.6$\pm$2.3 \\
               & 56000$-$56200      &22.1$\pm$2.5 \\
               & 56200$-$56400      &14.6$\pm$2.4 \\
               & 56400$-$56600      &27.9$\pm$2.7 \\
               & 56600$-$56800      &30.7$\pm$2.8 \\
               & 56800$-$57000      &24.5$\pm$3.4 \\
               & 57000$-$57200      &38.9$\pm$4.5 \\
               & 57200$-$57400      &18.3$\pm$2.3   \\
               & 57400$-$57600      &20.5$\pm$2.4   \\
               & 57600$-$57800      &14.9$\pm$2.1   \\
               & 57800$-$58000      &16.5$\pm$4.2   \\
\hline
  J1913$+$1330 & 55150$-$57400$^\dagger$  &4.7$\pm$0.2 \\
               & 55200$-$55400        &2.5$\pm$0.6 \\
               & 55400$-$55500        & 2.4$\pm$0.5\\
               & 55500$-$55750        &2.3$\pm$0.5 \\
               & 55750$-$55900        &2.1$\pm$0.5 \\
               & 55900$-$56170  & 6.6$\pm$0.6\\
               & 56170$-$56400        &6.8$\pm$0.7 \\
               & 56400$-$56600        &4.6$\pm$0.5 \\
               & 56700$-$56900        &10.9$\pm$1.2 \\
               & 56900$-$57100        &3.6$\pm$0.6 \\
               & 57100$-$57400        &4.8$\pm$0.6 \\
               & 57400$-$57600        &5.4$\pm$1.3 \\
               & 57600$-$57800        &6.6$\pm$1.3 \\
               & 57800$-$58000        &3.9$\pm$0.8 \\
\hline
  J1913$+$1330 & 55942$-$58000        &64.4$\pm$0.7$^\ddagger$\\
  (weak mode)  &                      &          \\ 
\hline
\end{tabular}
\\
$\dagger$ for the full range of observations\\
$\ddagger$ pulse emission rate considering detection of $\sim$ 7000 pulses in weak mode over 110 hours of observations\\
\end{table}

\begin{figure}
\begin{center}
\includegraphics[angle=-90,width=0.5\textwidth]{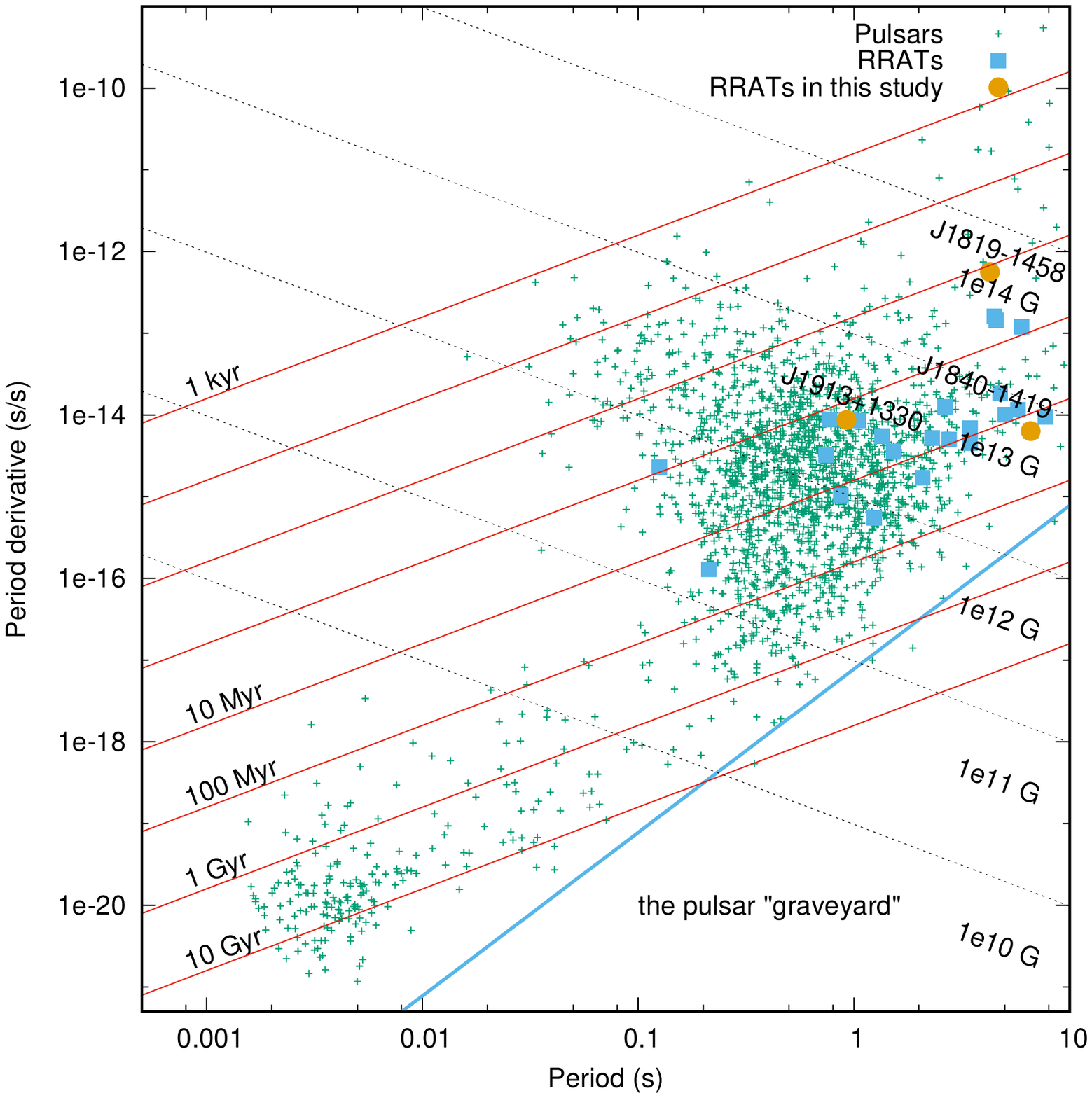}
\caption[ppdot_rrat]{The $P-\dot P$ diagram. RRATs studied in this paper, the other pulsars and RRATs for which timing solutions are available in the ATNF Catalogue \citep{manchester05} are presented. Lines of characteristic magnetic field (dotted lines), characteristic age (thin solid lines) and typical pulsar death line (\cite{lorimer}, thick solid line) are also indicated.}
\label{ppdot_rrats}
\vspace{0.1cm}
\end{center}
\end{figure}
\bsp    
\label{lastpage}
\end{document}